\documentclass[
    aps,
    pra,
    twocolumn,
    superscriptaddress,
    longbibliography,
    nofootinbib
    ]{revtex4-2}

\usepackage[
    top=1.5cm,
    bottom=2cm,
    left=1.5cm,
    right=1.5cm,
    paper=a4paper
    ]{geometry}

\usepackage[english,french]{babel}

\usepackage{graphicx,bm,amsmath,eufrak,amssymb,url}
\usepackage[usenames,dvipsnames]{color}
\usepackage[bbgreekl]{mathbbol}
\usepackage{xcolor}
\usepackage{braket}
\usepackage{tabularx}
\usepackage{adjustbox}
\definecolor{darkblue}{rgb}{0, 0, 0.8}
\usepackage[
    colorlinks=true,
    breaklinks=true,
    linkcolor=darkblue,
    citecolor=darkblue,
    urlcolor=darkblue,
    unicode,
    psdextra
    ]{hyperref}

\usepackage{float}
\usepackage{pgf}
\usepackage[T1]{fontenc}
\usepackage{verbatim}
\usepackage{graphicx}
\usepackage[
    font=small,
    labelfont=it,
    justification=justified,
    format=plain
    ]{caption}
\usepackage{subcaption}
\usepackage{url}

\newcommand\pdfmath[1]{\texorpdfstring{$#1$}{#1}}

\usepackage{bbold}

\newcommand{\be}{\begin{equation}}
\newcommand{\ee}{\end{equation}}
\newcommand{\bea}{\begin{eqnarray}}
\newcommand{\eea}{\end{eqnarray}}
\newcommand{\ba}{\begin{align}\begin{split}}
\newcommand{\ea}{\end{split}\end{align}}

\begin{document}
\title{A blueprint for a Digital-Analog Variational Quantum Eigensolver using Rydberg atom arrays}

\author{Antoine Michel}
\email{antoine.michel@edf.fr}
\affiliation{Electricité de France, EDF Recherche et Développement, Département Matériaux et Mécanique des Composants, Les Renardières, F-77250 Moret sur Loing, France }
\affiliation{Université Paris-Saclay, Institut d’Optique Graduate School,
CNRS, Laboratoire Charles Fabry, F-91127 Palaiseau Cedex, France}
\author{Sebastian Grijalva}
\author{Loïc Henriet}
\affiliation{PASQAL, 7 rue Léonard de Vinci, F-91300 Massy, France}
\author{Christophe Domain}
\affiliation{Electricité de France, EDF Recherche et Développement, Département Matériaux et Mécanique des Composants, Les Renardières, F-77250 Moret sur Loing, France }
\author{Antoine Browaeys}
\affiliation{Université Paris-Saclay, Institut d’Optique Graduate School,
CNRS, Laboratoire Charles Fabry, F-91127 Palaiseau Cedex, France}

\selectlanguage{english}
\date{\today}

\begin{abstract}
We address the task of estimating the ground-state energy of Hamiltonians coming from chemistry. We study numerically the behavior of a digital-analog variational quantum eigensolver for the H$_2$, LiH and BeH$_2$ molecules, and we observe that one can estimate the energy to a few percent points of error leveraging on learning the atom register positions with respect to selected features of the molecular Hamiltonian and then an iterative pulse shaping optimization, where each step performs a derandomization energy estimation.
\end{abstract}

\maketitle

\section{Introduction}\label{sec:introduction}

Quantum simulation holds the promises to solve outstanding questions
in many-body physics, in particular finding the ground state of strongly interacting quantum systems \cite{georgescu_quantum_2014, mcclean_theory_2016}. 
The determination of the ground state energies of complex molecules, one of the main tasks in quantum chemistry, is therefore an example of application where quantum simulation could be of interest. Along this line, proof-of-principle demonstrations were obtained using photons \cite{Lanyon2010,Peruzzo2014}, ions \cite{Shen2017,Shen2018,Hempel2018} or quantum circuits \cite{Kandala2017}.   
The last two examples used an hybrid approach were a classical computer optimizes in an iterative way the results obtained by a quantum device that was operating in a digital mode, i.e. as a series of one and two-qubit gates. 


Rydberg quantum simulators are another example of promising quantum simulation platforms thanks to their potential for scaling the number of qubits and their programmability \cite{browaeys_many-body_2020}. 
They rely on individual atoms 
trapped in arrays of optical tweezers that can interact when promoted to Rydberg states. The platform naturally implements spin Hamiltonians. Analog quantum simulation with hundreds of atoms has now been achieved \cite{scholl_programmable_2021, ebadi_quantum_2022,chen_continuous_2023}. 

One appealing  feature of this platform is the ability to place the atoms in arbitrary position in two and three dimensions, thus allowing large flexibility in their connectivity. Another feature  is their ability to prepare different initial product states as heuristic trials before the unitary evolution (whether it is by a set of digital gates or the action of an analog Hamiltonian evolution). However, this freedom in register preparation has a significant time cost that adds to the repetition clock rate \cite{henriet_quantum_2020}.

\begin{figure}[ht!]
    \centering
  \includegraphics[width=\linewidth,
                   trim={3.9cm 3cm 6cm 3cm},
                   clip]{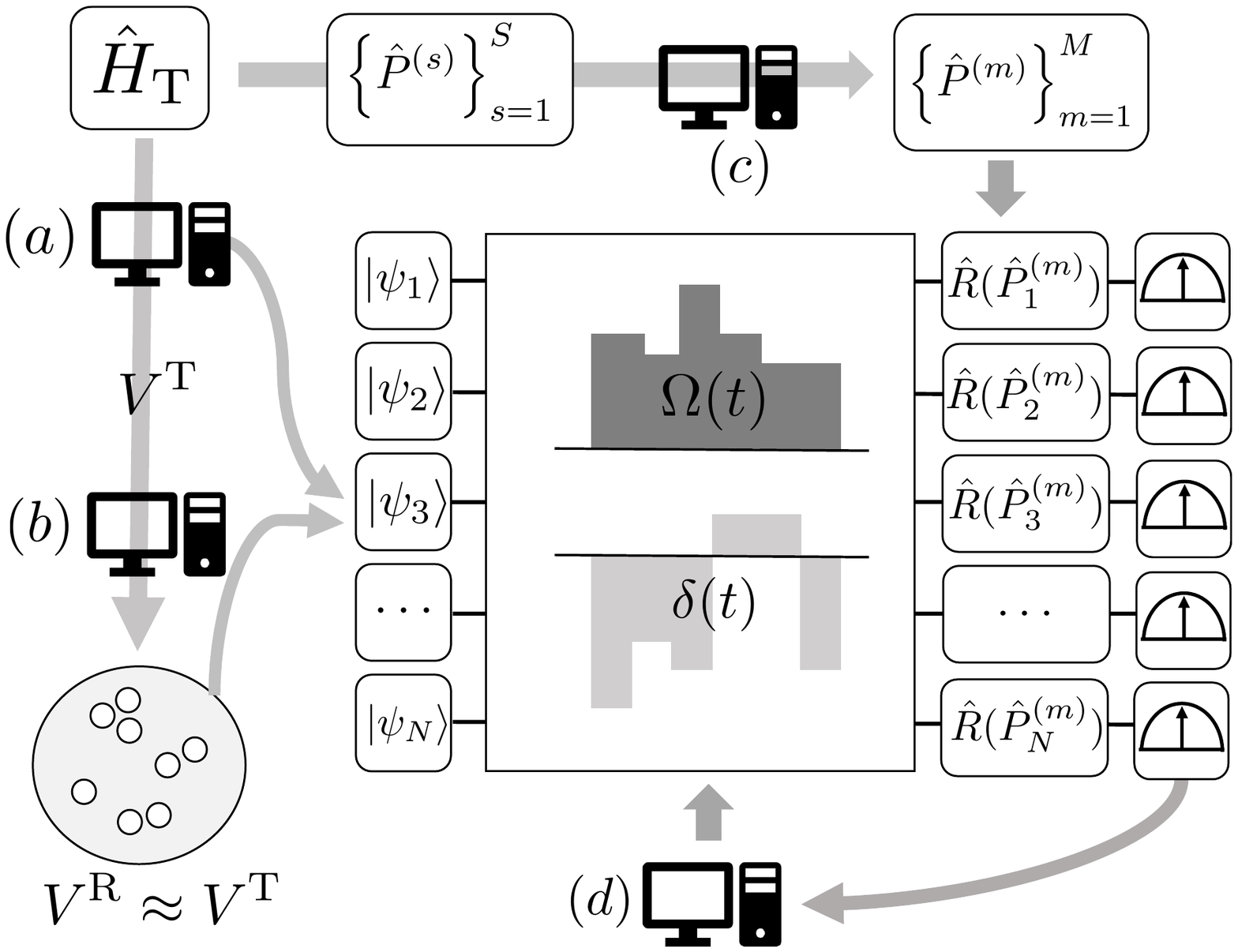}
\caption{Digital-Analog VQE on Rydberg atoms studied in this paper. The computer icons indicate where classical processing enhances and informs the cycle. First, a target molecular Hamiltonian $\hat H_\text{T}$ is passed to computer \emph{(a)}. This computer extracts a subset of terms that define a target interaction matrix $V^\text{T}$. Computer \emph{(b)} then optimizes atom positions (with interaction matrix $V^\text{R}$) to approximate $V^\text{T}$. With a chosen register, computer \emph{(a)} tests virtually some product states that can be experimentally implemented, warm-starting the algorithm. Meanwhile, computer \emph{(c)} takes the Pauli strings $\{ \hat P ^{(s)}\}$ that constitute $\hat H_\text{T}$ and outputs a \emph{derandomized} set of Pauli measurement basis $\{ \hat R(\hat P ^{(m)})\}$ that will be used at the end of the circuit where the prepared state is measured in the $Z$-axis. A parameterized pulse acts on the initial state, and the readout data is used by computer \emph{(d)} to estimate the expectation value of the target Hamiltonian $\langle \hat H_{\text{T}}\rangle$. Finally, computer \emph{(d)} calculates new pulse parameters to update the quantum evolution. When a desired precision or some stopping criteria is reached, the best energy value is returned.}
    \label{fig:VQE_analog}
\end{figure}
Neutral atom devices are naturally suited for \emph{analog quantum algorithms}, where the analog blocks are represented by control pulses that drive the system (or subsets of it). Given a prepared state, the parameterized pulses can be adjusted to variationally improve on a given score of the state. Methods for the optimization of parameters have been the subject of intense exploration in recent years \cite{wecker_progress_2015, cerezo_variational_2021, mcclean_theory_2016,barkoutsos_improving_2020,mcclean_barren_2018,meitei_gate-free_2021, wakaura_evaluation_2021, banchi_measuring_2021, gacon2021simultaneous, piskor_using_2022}. Additionally, the information and ``cost functions'' from the prepared quantum system are obtained by repeatedly measuring the state in the computational basis, which constitutes an operational overhead. Recent results \cite{huang_efficient_2021, elben_statistical_2019,kokail_self-verifying_2019,nam_ground-state_2020,ebadi_quantum_2022,dalyac_qualifying_2021} on protocols for the estimation of quantum observables are available and have helped establishing efficient measurement procedures based on generalized random measurements and a series of post-processing steps that are performed on a classical computer and that alleviate the measurement overhead. The types of randomized measurements that we shall describe in this paper require local rotations on the qubits of the register, thus constituting another ``digital'' layer, from a quantum circuit perspective. In fact, \emph{digital-analog} algorithms \cite{parra-rodriguez_digital-analog_2020}, benefit from the fact that analog operations can be performed with much higher fidelities than when using digital gates, while local single-qubit gates can be added explicitly in crucial steps of the process (state preparation and measurement). 

In this paper, we explore the implementation of a digital-analog VQE algorithm in a Rydberg quantum simulator. We account for typical constraints of the platform: the local action is restrained to the initial state preparation and measurement, with Hamiltonian time-evolution acting on the entire system as a ``global'' gate. We study numerically this version of a VQE for the $\text{H}_{\text{2}}$ molecule using common ansatze, followed by a more efficient protocol for larger molecules. We discuss the embedding of the Hamiltonian in the atom register, the way in which the optimization of the pulse sequence can be performed and the necessity of including an efficient estimation of energies (namely, we explore the effect of a \emph{derandomization} estimation \cite{huang_efficient_2021}) at each iteration step. We apply this numerically to the examples of LiH and BeH$_2$.
The manuscript is organized as follows: In section II we recall how the Variational Quantum Eigensolver (VQE) estimates the energy of the ground state of a molecular-based Hamiltonian. We then describe the basic ingredients of Rydberg Atom Quantum Processors and the Hamiltonians that they implement. We end the section by explaining the optimization cycle of variational quantum algorithms on these devices. In section III we describe the strategies for implementation of the VQE, going from a direct application of a Unitary Couple Cluster Ansatz, to the Quantum Alternating Operator ansatz and finally to a more hardware-oriented approach that combines elements of register preparation, pulse optimization and observable estimation. This is followed in section IV by numerical results of the error in energy obtained as a function of the number of repetitions of the experiment, an informative measure of the performance of hybrid classical-quantum implementations.

\section{Analog Variational Quantum Eigensolver with Rydberg atoms}
The Variational Quantum Eigensolver (or VQE) is a hybrid quantum-classical algorithm designed to find the lowest eigenvalue of a given Hamiltonian \cite{fedorov_vqe_2022}. We describe below the origin of the Hamiltonians that we consider and how VQE can be studied with a Rydberg Quantum Processor.

\subsection{Hamiltonians from Quantum Computational Chemistry}\label{sec:chem}

We first recall the method used to express the electronic Hamiltonian 
as a spin model (see e.g.~\cite{Hempel2018}). We start from the Born-Oppenheimer approximation of the Hamiltonian of the system, which considers the nuclei of the molecules as classical point charges:

\begin{equation}\label{elec-ham}
    \hat H = -\sum_i \frac{\nabla_i^2}{2} - \sum_{i, I} \frac{\mathcal Z_I}{|\mathbf r_i - \mathbf R_I|} + \frac 1 2 \sum_{i\neq j} \frac{1}{|\mathbf r_i - \mathbf r_j|} 
\end{equation}
(in atomic units) where $\nabla_i$ is the kinetic energy term for the $i$-th electron, $\mathcal Z_I$ is the charge of the $I$-th nucleus, and $\mathbf r$, $\mathbf R$ denote the distance of the $i$-th electron and the $I$-th nucleus with respect to the center of mass, respectively. We aim to obtain the ground state energy of  \eqref{elec-ham}.

One needs to define a basis set in which to represent the electronic wavefunctions. We shall concentrate on the Slater-type orbital approximation for the basis set, with three Gaussian functions, STO-3G. This minimal basis set $\{\phi_i (\mathbf x_i)\}$ (where $\mathbf x_i=(\mathbf r_i, \sigma_i)$ encodes the $i$-th electron's spatial and spin coordinates) includes the necessary orbitals to represent the valence shell of an atom. Moreover, the wavefunctions need to be anti-symmetric under the exchange of electrons. This can be achieved through \emph{second quantization}, where one defines anticommuting fermionic creation/annihilation operators $\{\hat a_p^\dagger\}, \{\hat a_p\}$ and rewrites the initial Slater determinant form of the wavefunction as $|\Psi \rangle = \prod_p (\hat a_p^\dagger)^{\phi_p} |\mathrm{vacuum}\rangle$, representing the occupation of each molecular orbital.

The fermionic operators are used to rewrite \eqref{elec-ham} as: 
\be
\label{eq:2nd-quant-ham}
 \hat H = \sum_{p,q}h_{pq}\hat{a}_p^{\dagger}\hat{a}_q + \frac{1}{2}\sum_{p,q,r,s}h_{pqrs}\hat{a}_p^{\dagger}\hat{a}_q^{\dagger}\hat{a}_r\hat{a}_s.
\ee
where the coefficients $h_{pq}$ and $h_{pqrs}$ encode the spatial and spin configuration of each of the electrons and depend on the inter-nuclear and inter-electron distances $\mathbf R, \mathbf r$: 
\be\label{eq:2nd-quant-terms}
\begin{aligned}
    h_{pq}&=\int d \mathbf x \phi_p^* (\mathbf x)\left(-\frac{\nabla^2}{2}-\sum_i\frac{\mathcal Z_i}{|\mathbf R_i- \mathbf r|}\right)\phi_q(\mathbf x) \\
    h_{pqrs}&=\int d\mathbf x_1 d\mathbf x_2\frac{\phi_p^*(\mathbf x_1)\phi_p^*(\mathbf x_2)\phi_r(\mathbf x_1)\phi_s(\mathbf x_2)}{|\mathbf r_1- \mathbf r_2|}.
\end{aligned}
\ee
Next, we map the fermionic operators acting on Fock states of $n$ orbitals to a Hilbert space of operators acting on spin states of $N$ qubits. This corresponds to the quantum processors' effective interaction Hamiltonians, quantum gates and measurement basis. Useful maps of this kind include the Jordan-Wigner (JW) \cite{jordan_uber_1928} or the Bravyi-Kitaev (BK) \cite{bravyi_fermionic_2002} transformations. The obtained Hamiltonian is a sum of tensor products of single-qubit Pauli matrices:
\begin{align}\label{eq:pauli_ham}
    \hat{H}_T = \sum_{s=1}^S {\bf c}_s \bigg( \bigotimes_{j=1}^N \hat{P}_j^{(s)}\bigg)
\end{align}
where $\hat P_j \in \{\mathbb 1, X, Y, Z \}$,  $S$ is the number of Pauli strings in the Hamiltonian and $N$ the number of qubits.



\subsection{Rydberg Atom Quantum Processor}
Rydberg atom arrays are now well-established quantum simulation 
platforms \cite{henriet_quantum_2020,browaeys_many-body_2020}. Briefly, atoms are trapped in optical tweezers, each
containing exactly one atom. The tweezers may be arranged in any 1D, 2D or 3D geometrical configurations. The register can be rebuilt after each computational cycle.
To perform quantum processing, we use the fact that the platform implements spin-like Hamiltonians, where the interactions originate from strong dipole-dipole couplings between atoms laser-excited to Rydberg states. 

Depending on the choice of atomic levels, the atoms experience different effective interactions. In the case of the Ising mode, $\ket{0}$ is a ``ground'' state prepared by optical pumping \cite{browaeys_many-body_2020} and $\ket{1}$ is a Rydberg state of the atom. The Hamiltonian term for this interaction is:

\begin{align}\label{eq:ising}
    \hat H_{\text{Ising}} &= \sum_{i > j}\frac{C_6}{r_{ij}^{6}}\hat n_i \hat n_j, 
   \end{align}
with $\hat n_i = |1\rangle_i \langle 1 | = (\mathbb 1_i + Z_i)/2$ the projector on the Rydberg state and $r_{ij}$ the distance between atoms. Here and below, $X_i,Y_i$ and $Z_i$ indicate the local Pauli operators.

If instead the two states chosen are two dipole-coupled Rydberg states (for example $\ket{0}=\ket{nS}$ and $\ket{1}=\ket{nP}$ for large $n$), the interaction is resonant and realizes 
a so-called ``XY'' or ``flip-flop'' term:

\begin{align}\label{eq:XY}
    \hat H_{\text{XY}} &= \sum_{i \neq j}\frac{C_3}{r_{ij}^{3}} (X_i X_j + Y_i Y_j),
\end{align}
where $C_3$ depends on the chosen Rydberg orbitals and their orientation with respect to the interatomic axis. It corresponds to a coherent exchange of neighboring spin states $\ket{10}$ to $\ket{01}$.

In addition, we can include time-dependent terms on the Hamiltonian, by means of a laser pulse (Ising mode) or a microwave field (XY mode) targeting the transition between the ground and excited states. This is represented by the following ``drive'' terms:

\begin{align}
\hat H_{\text{drive}} = \frac{\hbar}{2}\sum_{i=1}^N \Omega_i(t) X_i - \hbar\sum_{i=1}^N \delta_i(t)\hat n_i\ .
\end{align}
Here, $\Omega(t)$ is Rabi frequency and  $\delta(t)$ the detuning of the field  with respect to the resonant transition frequency. 
The addressing can be either global or local. In the procedure used in this work, the 
local addressing is restricted to the initial state preparation and the register readout stages. 

\subsection{Variational Algorithms on a Rydberg atoms device}

In the analog VQE algorithm, we seek to estimate the energy of the ground state of a qubit Hamiltonian called the \emph{target} Hamiltonian, $\hat H_{\text{T}}$,  by using an iterative method. 
The \emph{resource} Hamiltonian is the one realized by the hardware, 
and can be configured with different types of interactions ($\hat H_\mathrm{inter}$) (\ref{eq:ising}, \ref{eq:XY}) and driving fields ($\hat H_\mathrm{drive}$): 
\begin{equation}
    \hat H_{\text R} = \hat H_{\rm inter} + \hat H_{\rm drive}.
\end{equation}
Experimentally, the transition from the ground to the excited state is typically generated by a two-photon process, from which an approximate two level system is extracted, driven by an effective Rabi frequency $\Omega$ and detuning $\delta$ during the quantum processing stage. We use their values as parameters in our analog presentation of a VQE algorithm: The first step is to prepare the register of $N$ atoms with a geometry that determines the interaction terms $\hat H_{\rm inter}$ and then to initialize the system in a state $\ket{\psi_0}$. 
Then, a pulse sequence is applied to evolve the system under the resource Hamiltonian $\hat H_{\text R}(\Omega(t), \delta(t))$ whose corresponding time-ordered unitary evolution operator is $U(t) = \mathcal T  \exp\big(-i \int_{0}^{t}\hat H_{\text{R}}(\Omega(\tau), \delta(\tau) )d\tau\big)$. The final prepared state is:
\be
\label{eq:evolution_op}
\ket{\psi(\Omega, \delta,t)} = U(t)\ket{\psi_0}.
\ee
The energy of a prepared state will be calculated with respect to the target Hamiltonian: 
\be
\label{eq:energy_estimation}
E(\Omega, \delta, t) = \bra{\psi(\Omega, \delta, t)}\hat H_{\text T}\ket{\psi(\Omega, \delta, t)}.
\ee
After each cycle, a classical optimizer adjusts the parameters $\Omega \rightarrow \Omega'$, $\delta \rightarrow \delta'$ and $t \rightarrow t'$ and we repeat the evolution of the initial quantum state $\ket{\psi_0}$ with the new parameter set $U(\Omega',\delta',t')$. We aim to obtain for each iteration $E(\Omega', \delta', t') \leq E(\Omega, \delta, t)$ \footnote{This classical optimization problem can be addressed for example by obtaining the gradient of the energy function.}. After several iterations of this loop, the variational scheme attempts to prepare a state whose energy is a good approximation of the ground state energy of $\hat H_{\text T}$ \cite{mcclean_theory_2016}.

\section{Description of the Protocols}

In this section, we describe two analog variational quantum algorithms for the estimation of the ground state energy and apply them to quantum chemistry problems. The protocols differ mainly by the choice of ansatz: one is the \emph{Unitary Coupled Cluster} (UCC) ansatz \cite{bartlett_alternative_1989}, while the other is an adaptation of a \emph{hardware-efficient ansatz} \cite{Kandala2017}, based on repeating alternating values of amplitude, frequency or phase of the applied pulses. We verify numerically the performance of these two types of ansatz in a Rydberg-based Quantum Processor (QP). Next we discuss a protocol for larger molecules tailored after the hardware capabilities. We begin by considering the prototypical example of the H$_2$ molecule.

\subsection{UCC ansatz on an analog quantum processor: application on H\pdfmath{\bf _2}}

Numerous implementations of the VQE algorithm rely on the use of  digital gates. 
Recent experimental implementations for the H$_2$, LiH and BeH$_2$ molecules have been realized in \cite{Kandala2017, Hempel2018}, with superconducting and trapped ions devices respectively. For the analog version of this algorithm on H$_2$, we consider the target Hamiltonian and the ansatz as in \cite{Hempel2018}.
The Jordan-Wigner and Bravyi-Kitaev transformations lead to two different spin Hamiltonians of this molecule:

\begin{align}
\begin{split}
\hat{H}_{\text{JW}} =& \, {\bf c_0} \mathbb{1} + {\bf c_1}(Z_0 +  Z_1) + {\bf c_2}( Z_2 +  Z_3) +\\
&{\bf c_3} Z_3 Z_2 +  {\bf c_4} Z_2 Z_0 + {\bf c_5}(Z_2 Z _0 + Z_3 Z_1) +\\
&{\bf c_6} (Z_2 Z_1 + Z_3 Z_0) + {\bf c_7} (X_3 Y_2 Y_1 X_0 +\\
& Y_3 X_2 X_1 Y_0 - X_3 X_2 Y_1 Y_0 + Y_3 Y_2 X_1 X_0)
\label{hamjw}
\end{split}
\end{align}    
and
\begin{align}
\begin{split}
\hat{H}_{\text{BK}} =& \, {\bf f_0} \mathbb{1}  + {\bf f_1} Z_0 + {\bf f_2} Z_1 + {\bf f_3} Z_2 + {\bf f_4}Z_1 Z_0 +\\ & {\bf f_5} Z_2 Z_0 + {\bf f_6} Z_3 Z_1 +  {\bf f_7} X_2 Z_1 X_0 +
    {\bf f_8} Y_2 Z_1 Y_0 + \\ 
    & {\bf f_9} Z_2 Z_1 Z_0 + {\bf f_{10}} Z_3 Z_2 Z_0 + {\bf f_{11}}Z_3 Z_2 Z_1 + \\
    & {\bf f_{12}}Z_3 X_2 Z_1 X_0 + {\bf f_{13}} Z_3 Y_2 Z_1 Y_0 + {\bf f_{14}}Z_3 Z_2 Z_1 Z_0
\label{hambk}
\end{split}
\end{align}     
where the coefficients $\{\bf c_i\}$ and $\{\bf f_j\}$ are calculated from (\ref{eq:2nd-quant-terms}). Since in \eqref{hambk} qubits $1$ and $3$ are only affected by the operators $\mathbb{1}$ and $Z$ one can actually work with the following two-qubit effective Hamiltonian \cite{omalley_scalable_2016}:
\begin{align}\begin{split}
\hat{H}_{\text{BK}}^{\text{(eff)}}=& \, {\bf g_0} \mathbb{1} + {\bf g_1}Z_0 + {\bf g_2}Z_1 + {\bf g_3} Z_0 Z_1 + \\
& {\bf g_4}X_0 X_1 + {\bf g_5}Y_0Y_1.
\label{eqhambkef}
\end{split}\end{align}

Usually, a good ansatz $|\psi(\boldsymbol \theta)\rangle = U(\boldsymbol \theta) |\psi_0\rangle$ requires a balance between hardware constraints and symmetries in target Hamiltonian. However, using the `knobs' available on the hardware is often not efficient, and one thus needs additional guidance to reach the states we are looking for in a potentially very large Hilbert space. In this sense, the well-established Unitary Coupled Cluster (UCC) ansatz allows one to perform an unitary operation while keeping advantages of coupled cluster ansatz from chemistry \cite{helgaker_molecular_2014}. 

In most cases, implementing the UCC ansatz in a quantum processor requires constructing a digital quantum circuit with full local addressing. An example where global addressing is sufficient is the H$_2$ molecule. The initial guess of the molecular wave function is a product state obtained from the classical Hartree-Fock calculation performed to determine the coefficients $\{\bf c_i\}$ and $\{\bf f_j\}$. Considering only relevant single and double excitations in the unitary coupled-cluster operator (UCC-SD) yields the following one-parameter unitary: 
\begin{equation}\label{UCCSD}
    U_{\text{UCC-SD}}(\theta) = e^{\theta (c_{2}^{\dagger}c_{3}^{\dagger}c_1c_0 - c_0^{\dagger}c_1^{\dagger}c_3c_2)}
\end{equation}
where the minimal set of orbitals are represented by the fermionic annihilation and creation operators  $c, c^{\dagger}$ \cite{Hempel2018}. A Jordan-Wigner transformation on these operators leads to the UCC ansatz $\ket{\psi(\theta)} =\exp(-i\theta X_3 X_2 X_1 Y_0) \ket{0011}$, where $\ket{0011}$ is the Hartree-Fock state. In the case of the (effective) Bravyi-Kitaev transform \eqref{eqhambkef}, we obtain the simpler UCC ansatz $\ket{\psi(\theta)}=\exp(-i\theta X_1 Y_0)\ket{01}$.
 
Since the evolution Hamiltonian commutes with the XY Hamiltonian (\ref{eq:XY}), one can use the latter ansatz and attempt to drive the Rydberg QP in the XY mode, 
using $\Omega=0$ and non-zero local detunings, leaving the rest of the parameters to be set by variational optimization:
\begin{align}
\begin{split}
\label{eq:xyansatz}
\ket{\psi(\delta_0, \delta_1, t)} &= \exp \Big( -it(\delta_0 Z_0 + \delta_1 Z_1 + \hat H_{\text{XY}}) \Big) \ket{01} \\ 
&= a(\delta_0,\delta_1,t) \ket{01} + b(\delta_0, \delta_1, t) \ket{10},
\end{split}
\end{align}
which coincides with the subspace reached with the UCC ansatz: 
\begin{equation}\label{eq:uccbk}
\exp(-i\theta X_1 Y_0)\ket{01} = a(\theta) \ket{01} + b(\theta) \ket{10}.
\end{equation}

\begin{figure}
    \centering
    \includegraphics[width=\linewidth]{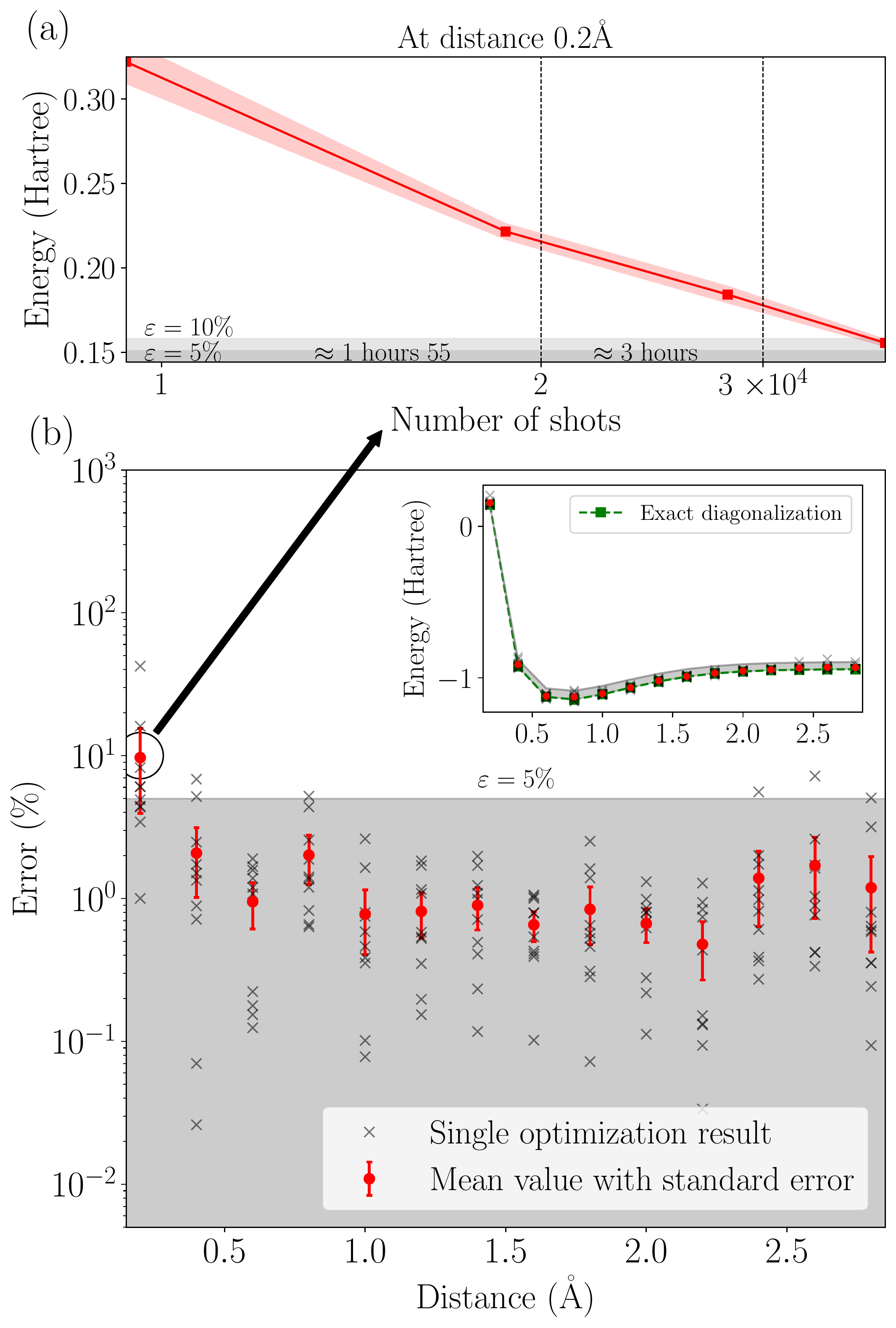}
    \caption{\textit{Numerical implementation of an analog VQE algorithm using a UCC ansatz.} 
    (a) A zoom on the smallest inter-atomic distance ($0.2$ \r{A}) shows the evolution of the optimization with respect to the number of shots. The differential evolution was set to perform at most 4 iterations (red squares). The red scale shows the errorbar over 20 realizations. It takes approximately $3.5$ hours of runtime for a QP operating at 3 Hz 
    to achieve $\epsilon = 10 \% $ (light grey scale) of error and $4$ hours to achieve $\epsilon = 5 \% $ (dark grey scale).
    (b) Relative error in percentage (red circles) between the mean VQE result and the numerically computed lowest eigenvalue of the target Hamiltonian (in STO-3G basis) over 10 realizations (gray crosses). The expected error is below $\varepsilon = 5 \%$ (grey area). The inset depicts the same result on an energy scale and compares it with the exact solution (green squares). The result is drawn as a function of hydrogen inter-atomic distance. 
    }
    \label{fig:H2_UCC_plot}
\end{figure}

A numerical implementation of this protocol is shown in Fig. \ref{fig:H2_UCC_plot}, where the classical optimization was performed with a differential evolution algorithm \cite{storn_differential_1997}. We observe that the ground-state energy can be obtained with an error smaller than $5\%$ using less than $36500$ shots for each point. 

Such examples of a UCC ansatz implementable with an analog approach, often rely on finding symmetries between target and resource Hamiltonians \cite{kokail_self-verifying_2019}. Nevertheless, this kind of protocol remains impractical for larger molecules because of the increasingly higher number of qubits and Pauli strings in the Hamiltonian. In order to use the analog approach for larger encodings, we explore other approaches below.

\subsection{Alternating pulses}

\begin{figure}[ht!]
    \centering
    \includegraphics[width=1.1\linewidth,trim={0cm 0cm -1cm 0.5cm},clip]{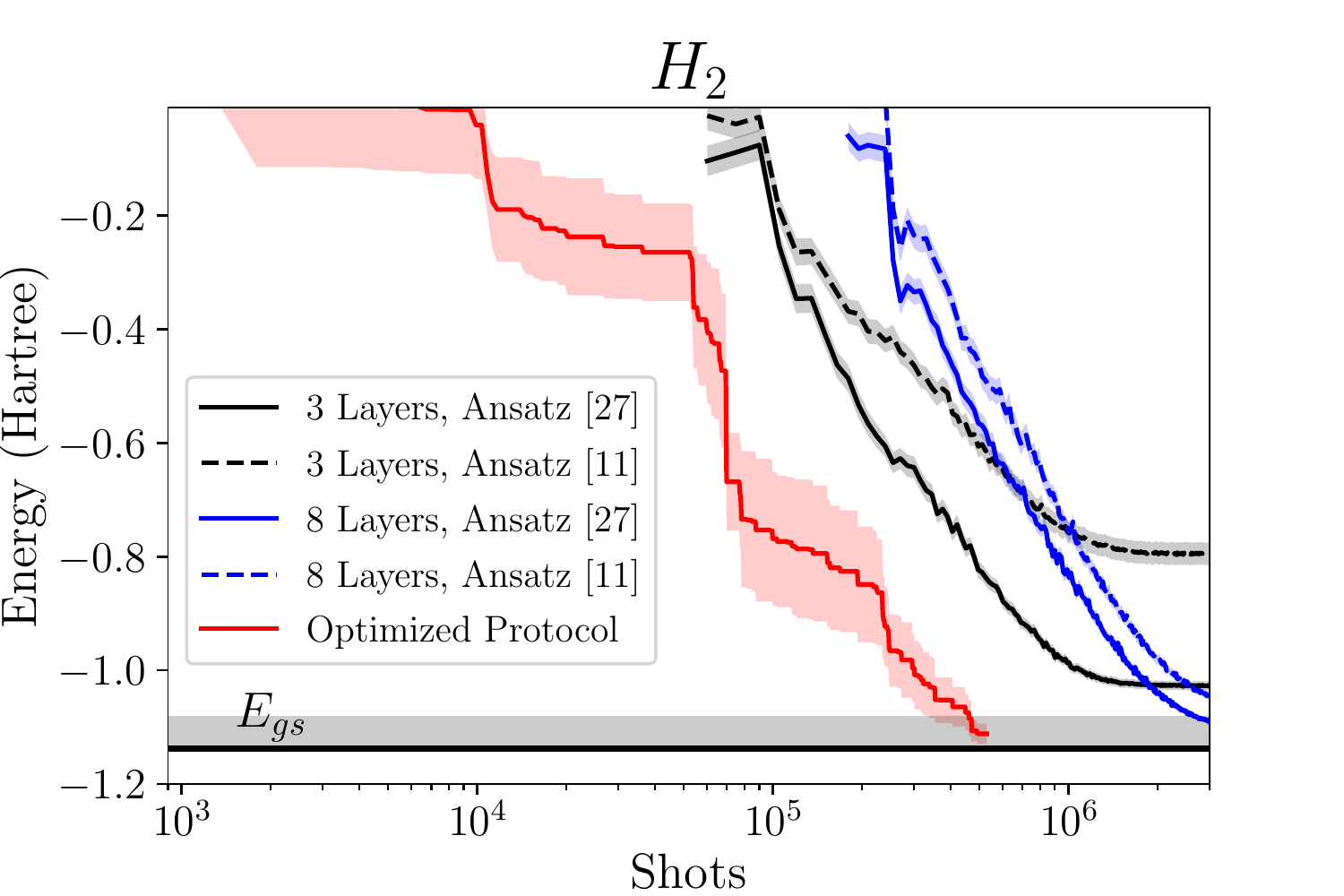}
    \caption{Evolution of the ground-state energy as a function of the accumulated number of shots for the H$_2$ molecule at a fixed inter-nuclear distance. We have averaged numerically 200 realizations of VQE with 3 and 8 layers using the two alternating pulse ansatz, \cite{dalyac_qualifying_2021} (straight line), and \cite{ebadi_quantum_2022} (dashed line). In the alternating operator approach, each Pauli string mean value is performed with 1000 shots, which for the H$_2$ Hamiltonian represents $1.5\times10^4$ shots before obtaining the first energy data point. Achieving energy errors below $5\%$ (gray area) requires at least $\mathcal O (10^6)$ shots in total. For the optimized procedure (adding more control over atom positions, pulse shaping and derandomization estimation), the same energy error typically requires $\mathcal O(10^5)$ shots.} \label{fig:QAOA_comp}
\end{figure}

We now describe an alternating operator approach, based on the QAOA algorithm \cite{farhi_quantum_2014}. Let $|\psi_0\rangle$ be the state composed of all qubits in the ground state. The whole sequence is composed by alternating constant (global) pulses, corresponding to two non-commuting Hamiltonians $\hat H_a, \hat H_b$:
\begin{align}
    \hat H_{a} &= \frac{\hbar}{2}\sum_{i=1}^N \Big( \Omega X_i - \delta Z_i \Big) + \hat H_{\text{inter}} \\
    \hat H_{b} &= \frac{\hbar}{2}  \sum_{i=1}^N \Omega X_i  +\hat H_{\text {inter}}.
\end{align}
These Hamiltonians define evolution operators $U_a(t)$ and $U_b(t)$, during a certain time $t$ (see \eqref{eq:evolution_op}). The ansatz of $L$ layers is written as: 
\be
\ket{\psi(\mathbf t_a, \mathbf t_b)}= \prod_{\ell=1}^L U_a(t^{\ell}_a) U_b(t^{\ell}_b) \ket{\psi_0},
\ee
where the arrays of parameters $\mathbf t_k = (t^{1}_k, \ldots, t^{\ell}_k, \ldots, t^{L}_k)$, $k \in \{a,b\}$, fix the duration of each pulse in the layer, as described in \cite{dalyac_qualifying_2021}. 
As another example, a different choice of parameters was used in \cite{ebadi_quantum_2022}, considering a single Hamiltonian:
\begin{align}
\begin{split}
    \hat H = \frac{\hbar}{2}&\sum_{i=1}^N \Big( \Omega(t) e^{i\phi(t)} \ket{0}_i\bra{1} + \text{h.c.} \Big)  + \hat H_{\text{inter}}
\end{split}
\end{align}
with different time $\mathbf t = (t^{1}, \ldots, t^{\ell}, \ldots, t^{L})$ and phase $\boldsymbol \phi = (\phi^{1}, \ldots, \phi^{\ell}, \ldots, \phi^{L})$ arrays defining $L$ segments of the pulse. The corresponding ansatz is then:
%
\be
\ket{\psi(\mathbf t, \boldsymbol{\phi})}= \prod_{\ell=1}^L U(t^{\ell}, \phi^{\ell})  \ket{\psi_0}.
\ee
The two approaches can be implemented in existing experimental setups, especially when the target Hamiltonian is equal to to the resource Hamiltonian (such as the case of the Maximal Independent Set problem with Unit Disks, which is native to the Rydberg atoms setting). However, these methods struggle to minimize the molecular target Hamiltonian energies within a limited number of iterations and measurement repetitions. The alternating pulse ansatz assumes an initial register configuration and initial guesses for the durations of the pulses in each layer, two tasks that are the subject of active research. The expectation is that a properly chosen register and an optimized pulse will drive the system to a low-energy state. In Fig. \ref{fig:QAOA_comp}, we compare numerically the  performance of the two alternating pulse ansatze discussed above (\cite{dalyac_qualifying_2021}, \cite{ebadi_quantum_2022}). We also included the procedure described in Sec. \ref{Param_pulse}, which addresses the embedding of the problem in the register and an estimation protocol for the observables. Comparing the required number of shots for these approaches highlights the necessity of including an efficient estimation protocol for the observables.

\subsection{Optimized Register and Iteratively Parameterized Pulses}\label{Param_pulse}

In this section, we present a more refined approach to deal with larger systems,  aiming at exploiting the capabilities already available in Rydberg simulators. To exemplify the procedure, we consider in the following  the Ising mode with the  resource Hamiltonian (\ref{eq:ising}).

\subsubsection{Atom register and initial state}\label{sec:embedding}

Even though we only consider global pulses for the processing stage, there still remains freedom in the choice of the positions of the atoms. This determines the strength of pairwise interactions and defines a connectivity graph whose edges correspond to the atoms that experience a blockade effect \cite{henriet_quantum_2020} (a different graph structure can be defined for the XY mode \eqref{eq:XY}).

In order to find suitable atomic positions, the coordinates are optimized in the plane so that the associated interaction energy matrix resembles as much as possible the information contained in the target Hamiltonian. 
Since the latter contains  general Pauli strings, we consider a subset of terms whose coefficients can be expressed in terms of the coordinates of the atoms\footnote{A broader series of techniques for embedding the problem information into the atom register has been considered in \cite{leclerc2022financial, coelho2022efficient}}. 
A simple choice consists in selecting the terms that can be directly compared with the Ising-like interaction of the atoms: Let the matrix $V^{\text{T}}$ be given by the positive coefficients of the terms with only two $Z$ operators in the molecular target Hamiltonian and $V^{\text{R}}$ (our ``register'' matrix) the resulting values of interaction strength $C_6/r_{ij}^6$ for each pair $i,j$ of atom positions in the register. This defines a score function $\sum_{i,j}(V^{\text{T}}_{ij} - V^{\text{R}}_{ij})^2$ that we minimize numerically by varying the atom coordinates. 

\begin{figure}
    \centering
\includegraphics[width=0.9 \linewidth]{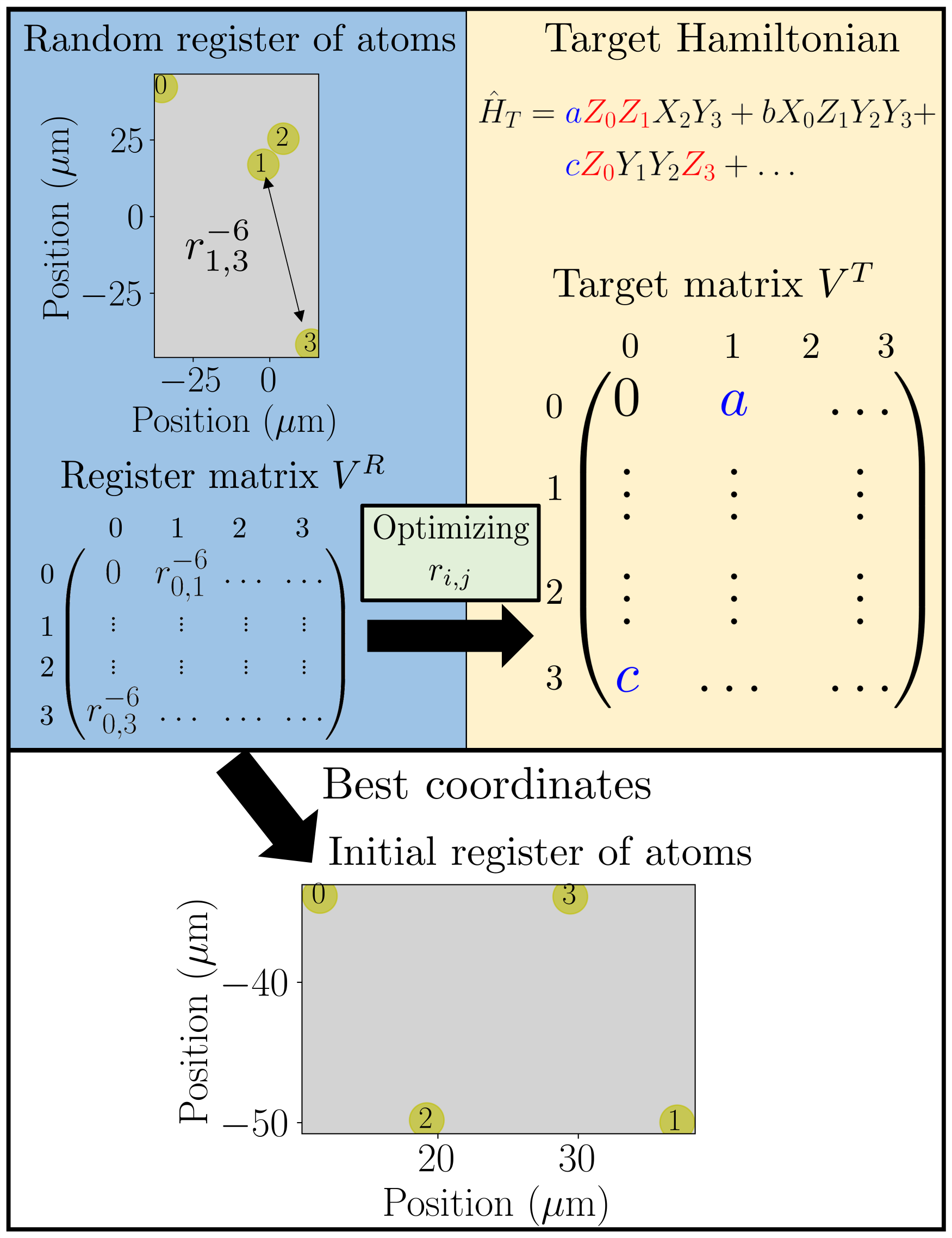}
    \caption{Protocol to optimize the positions of the atoms in a register based on a target Hamiltonian. Blue: we begin with a register of randomly placed atoms and all the Ising interaction terms are entered in a $N\times N$  matrix. Yellow: all positive coefficients before the Pauli strings with only two $Z$ operators in the target Hamiltonian are combined in another (target) matrix $N \times N$. The coordinates of the atoms are optimized to minimize the distance between the two matrices. We then obtain a new register on which we will apply the VQE sequence. }
    \label{fig:register_H2}
\end{figure}

The set of atomic positions that arises from this minimization will be our optimized register. Its geometry will be used to simulate the target Hamiltonian, but has no intrinsic chemical meaning. The information that is taken from the Hamiltonian can be chosen from other subsets of the Pauli strings (e.g. terms with 3 or more $Z$ operators) and different interpretations of how the coefficients constitute a register matrix. A different resource Hamiltonian, such as one with XY interactions, would imply a different choice of subset. In Fig. \ref{fig:register_H2}, we summarize graphically the procedure for the case of the $\text{H}_{\text{2}}$ molecule with the Jordan-Wigner transformation. It turns our that we obtain at a  geometry 
very similar to the one heuristically picked for the alternating pulse ansatz.

\subsubsection{Optimization of the parameterized pulse sequence}\label{sec:pulse_opt_protocol}

\begin{figure}[h!]
    \centering
    \includegraphics[width=\linewidth, trim=4cm 4.5cm 10cm 3.2cm, clip]{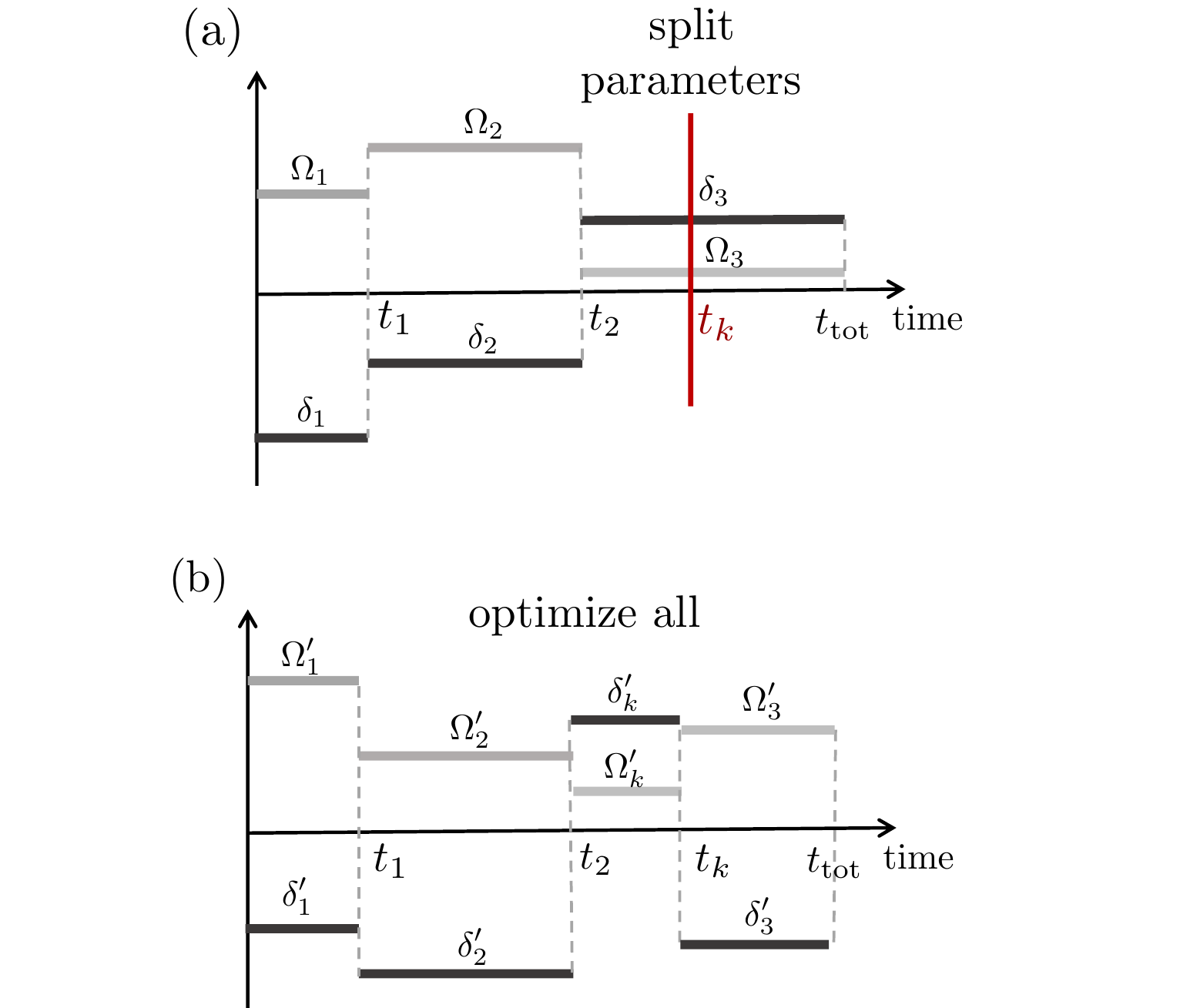}
    \caption{Iterative splitting and optimization of the pulse parameters: (a) Choose at random a time $t_{k}$, which will fall in the interval $(t_{i-1}, t_i)$, and accept it if $|t_{i-1}-t_{k}|$ and $|t_{k}-t_{i}|$ conform to the device response time. Split the corresponding $\Omega_{i}$ into two parameters $\Omega'_{i}, \Omega'_{k}$ with initial value equal to $\Omega_{i}$. Do the same to the set $\{ \delta_i \}$. (b) Optimize the new set of parameters to lower the energy of the prepared state (\ref{eq:energy_estimation}).}
    \label{fig:ctrl_pulse_sequence}
\end{figure}

We constructed a variation of the so-called \texttt{ctrl-VQE} protocol \cite{meitei_gate-free_2021} for the case of a global pulse on the register, in which the number of parameters increases at every optimization iteration, while the total time $t_\text{tot}$ remains fixed:

Consider a set of Rabi frequencies $\{\Omega_i\}_{i=1}^K$ and detunings $\{\delta_i\}_{i=1}^K$ defined discretely over a set of time labels $0 < t_1 < \ldots < t_K = t_\text{tot}$. 
Then, at iteration $k$, a new time label $0<t_{k}<t_\text{tot}$ is generated at random, lying between two previous time labels, $t_{i-1}< t_{k} < t_i$. To avoid labels too close to each other, we will accept $t_k$ if the intervals $|t_{i-1} - t_k|, |t_{k} - t_i|$ are large enough compared to the response time of the waveform generator of the machine (in the order of a few ns). The corresponding Rabi frequency $\Omega_i$ and detuning $\delta_i$ from the parent interval $[t_{i-1}, t_i]$ are then split into two independent parameters $\Omega_{i}', \Omega'_{k}$ and $\delta'_{i}, \delta'_{k}$ whose initial values are set equal to their parent parameters  (see Fig. \ref{fig:ctrl_pulse_sequence}). Finally, the new set of parameters $\{\Omega_1, \ldots, \Omega'_{i}, \Omega'_{k}, \ldots, \Omega_K \}$ (likewise for $\{\delta_i \}_{i=1}^K)$ is optimized starting from the previous iteration values. This algorithm acts therefore as a pulse shaping process. From time $t_{i-1}$ to $t_{i}$ the acting Hamiltonian is:
\begin{equation}
\hat H_i = \frac{\hbar}{2} \Big( \Omega_i \sum_{j=1}^N X_j - \delta_i \sum_{j=1}^N Z_j \Big) + \hat H_{\text{inter}}
\end{equation}
and our ansatz, for $K$ iterations, becomes:
\begin{equation}
    \ket{\psi(\boldsymbol \Omega,\boldsymbol \delta)} = \mathcal T \prod_{i=1}^K \exp\Big[-i \int_{t_i}^{t_{i+1}} \hat H_i(\tau) d\tau \Big ]\ket{\psi_0}
\end{equation}

Note that while the interval involves a constant Hamiltonian, we include a time-dependent integration at each interval, to indicate that the waveforms that compose the pulse can be adapted to hardware conditions (e.g. by being interpolated, or by adapting the shape with an envelope function).
\subsubsection{Energy estimation by derandomization}

In our algorithm implementation, we take as a figure of performance of the variational optimization run the \emph{total} number of shots required to achieve a given error threshold $\varepsilon$ for the energy. A bounded number of processing cycles is required to remain within a realistic time lapse for the entire implementation process. Rather than measuring several times each of the Pauli observables in the Hamiltonian, we use an estimation protocol (derandomization \cite{huang_efficient_2021}) based on fixing local Pauli measurements from an originally random set. This allows to efficiently predict the energy of the prepared state $\ket{\psi}$,  $\langle \hat H_\text{T}(\psi) \rangle$, at each loop of the optimization of the parameters. 

More specifically, the derandomization algorithm starts with an initial measurement set of $M$ random Pauli strings $\{\hat P^{(m)} \}_{m=1}^M$. A greedy algorithm improves the overall expected performance of the measurement set, effectively ``derandomizing'' the operators of each random Pauli string in sequence. The improvement is quantified by the average of the \emph{confidence bound}, which ensures that the empirical average \footnote{ 
A Pauli string $A$ \emph{hits} $B$, if by changing some operators in $A$ to $\mathbb 1$, we form $B$ (for example $ZX\mathbb{1}$ hits $\mathbb{1}X\mathbb{1}$ and $Z\mathbb{1}\mathbb{1}$). The empirical average is obtained using those Pauli measurement basis $\{\hat P^{(m)}\}$ that hit an observable $\hat P^{(s)}$, with the relevant measured bits expressed as $\pm$:
$$
\omega_s = \frac{1}{N_\mathrm{h}}\sum^M_{\substack{m: \\ \hat P^{(m)} \text{hits} \hat P^{(s)}}}  \Bigg(  \prod^N_{\substack{j : \hat P^{(s)}_j \neq \mathbb 1}} \mathbf b^{(m)}_j \Bigg),
$$
where $N_{\text{h}}$ counts how many Pauli strings in the set $\{\hat P^{(m)}\}$ hit $\hat P^{(s)}$, and $\mathbf b^{(m)} = \mathbf b^{(m)}_1 \cdots \mathbf b^{(m)}_N$ is the bitstring measured with the basis $\hat P^{(m)}$

}  $\omega_s$ corresponding to the $s$-th term of $\hat H_\text{T}$ is within a desired accuracy $| \omega_s - \langle \hat P^{(s)}\rangle |/|\langle \hat P^{(s)}\rangle | < \epsilon$ and with a high probability. The total energy is finally estimated as $\langle \hat H_\text{T}(\psi) \rangle\approx \sum_{s=1}^S \omega_s$.

While the pulses that prepare the state are global, the measurement itself requires the implementation of local rotations on the qubits. This can be achieved experimentally by using a toolbox such as the one described in \cite{notarnicola_randomized_2021}, thus emphasizing the digital-analog interplay that is now within reach for next-generation neutral atom devices.

\section{Numerical Results} \label{numerical_result}

\subsection{Application on LiH and BeH\pdfmath{\bf _2} molecules}\label{sec:result_lih_beh2}

\begin{figure}[ht!]
    \centering
    \includegraphics[width=1\linewidth]{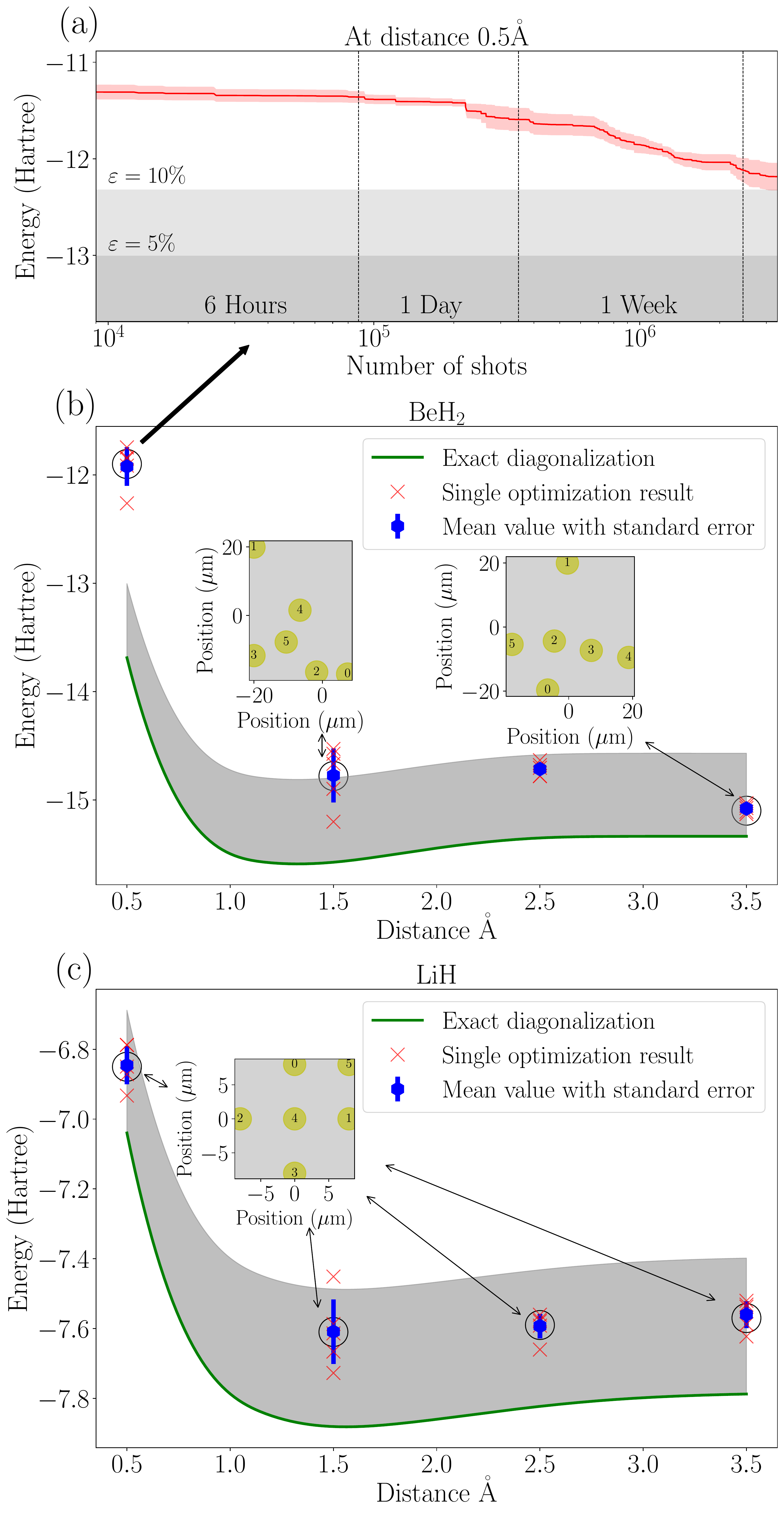} 
    \caption{\textit{Numerical results of the VQE algorithm with our digital-analog protocol} (a) for BeH$_2$ at an intermolecular distance of $0.5$ \r{A} where we increased the number of shots beyond 1 day of experiment and  observed the expected improvement over several days of calculations. The light grey shade and the dark grey shade indicate respectively $\varepsilon=10\%$ and $\varepsilon=5\%$ error benchmark. The red line shows the improvement mean value over 100 run.  
    (b) Result of BeH$_2$ molecule with an encoding of 6 qubits and 165 Pauli strings and (c) result of LiH molecule result with an encoding of 6 qubits and 118 Pauli strings. The insets show the register geometry at specific inter-nuclear distances. For the case of LiH a single heuristic choice was used, while for BeH$_2$ an optimized geometry was prepared at each inter-nuclear distance, minimizing the distance between selected terms of the target Hamiltonian and the interaction energies of the atoms in the register (see Sec. \ref{sec:embedding}). Blue squares:  mean value for several simulations.
    Green line: result from exact diagonalization. The gray shade indicates an $\varepsilon=5\%$ error benchmark. The total number of shots for each optimization result (red crosses) is set to 350000, corresponding roughly to a day of processing in a Rydberg QP.}
    \label{fig:results}
\end{figure}

We have applied the method described in section \ref{Param_pulse} to the LiH and BeH$_2$ molecules. Using the \texttt{Qiskit} \cite{Qiskit} framework combined with \texttt{Pyquante} \cite{muller_pyquante2_2022}, we calculate the one and two-body integrals of (\ref{eq:2nd-quant-terms}), encoding the problem into 6 qubits using the Bravyi-Kitaev method. The Hamiltonians contain 118 and 165 Pauli strings respectively. To design the pulse sequence and include realistic device constraints into the simulations we used the open source package \texttt{Pulser} \cite{silverio2022pulser}. The Powell algorithm \cite{powell_efficient_1964} was used for the classical optimization of the pulse values with 20 function evaluations for each iteration. The two initial Rabi frequency and detuning are chosen randomly in the interval $[0,2\times2\pi]$MHz for each optimization procedure. During the optimization, Rabi frequencies are bounded to this interval to remain within experimentally accessible values \cite{scholl_programmable_2021}, while the interval for the detuning was taken as $[-2\times 2\pi,2\times2\pi]$MHz. We ran the algorithm five times for four different inter-nuclear distances $R$ yielding the results shown in Fig. \ref{fig:results}. The algorithm converges with small errors in most cases, but we notice the impact of the initial parameters on the obtained energies. For instance at $R=1.5$ {\AA}  for the BeH$_2$ molecule, the obtained energy values are up to $0.4$ Hartree apart. 

To optimize the register geometry, we took the coordinates as parameters, starting from random positions  and minimized the score function described in Sec. \ref{sec:embedding}. We optimized the atom register based on the Nelder-Mead method \cite{nelder_simplex_1965}, with a few thousand function evaluations, and we also compared to several heuristic choices obtained by a term-by-term comparison with the interaction matrix (a well-performing choice of positions is shown for LiH in Fig. \ref{fig:results})

We define our error as
\be
\varepsilon = |{E_{\text{exact}} - E_{\text{estimated}}|/|E_{\text{exact}}}|,
\ee
where $E_{\text{exact}}=\langle \hat H_{\text{T}} \rangle$ is the exact diagonalization solution with respect to the target Hamiltonian and $E_{\text{estimated}}$ is the energy calculated with the optimized geometry, the optimized pulse sequence, and the derandomization estimation. The optimized  configurations, together with optimized pulse parameters and energies estimated at each iteration with derandomized measurements give rise to energy errors typically below the $\varepsilon = 5\%$ threshold in less than 350000 shots.

The implementation of the derandomization algorithm allows us to choose the number of measurements that we wish to take (our budget), for a given target accuracy of estimation, which we set to correspond to our $\varepsilon = 5\%$ benchmark. The resulting ``derandomized'' Pauli measurements included typically close to 20 different Pauli strings, calculated from the minimization of the average confidence bound that ensures an empirical average within the chosen $\varepsilon$. Since some derandomized Pauli strings have more operators in common with the terms in the target Hamiltonian (they ``hit'' more target observables), we adjusted the measurement repetitions to be spent proportionally more in them, which improved statistics. We also verified that the obtained accuracy improves upon increasing the allowed number of shots, although we don't expect a full convergence, given the incomplete information used to define $\hat H_{\text{T}}$.

\subsection{Roadmap for more complex molecules}

\begin{figure}[ht!]
    \centering
    \includegraphics[width=0.485\linewidth,trim={5.5cm 1.8cm 8.5cm 1cm},clip]{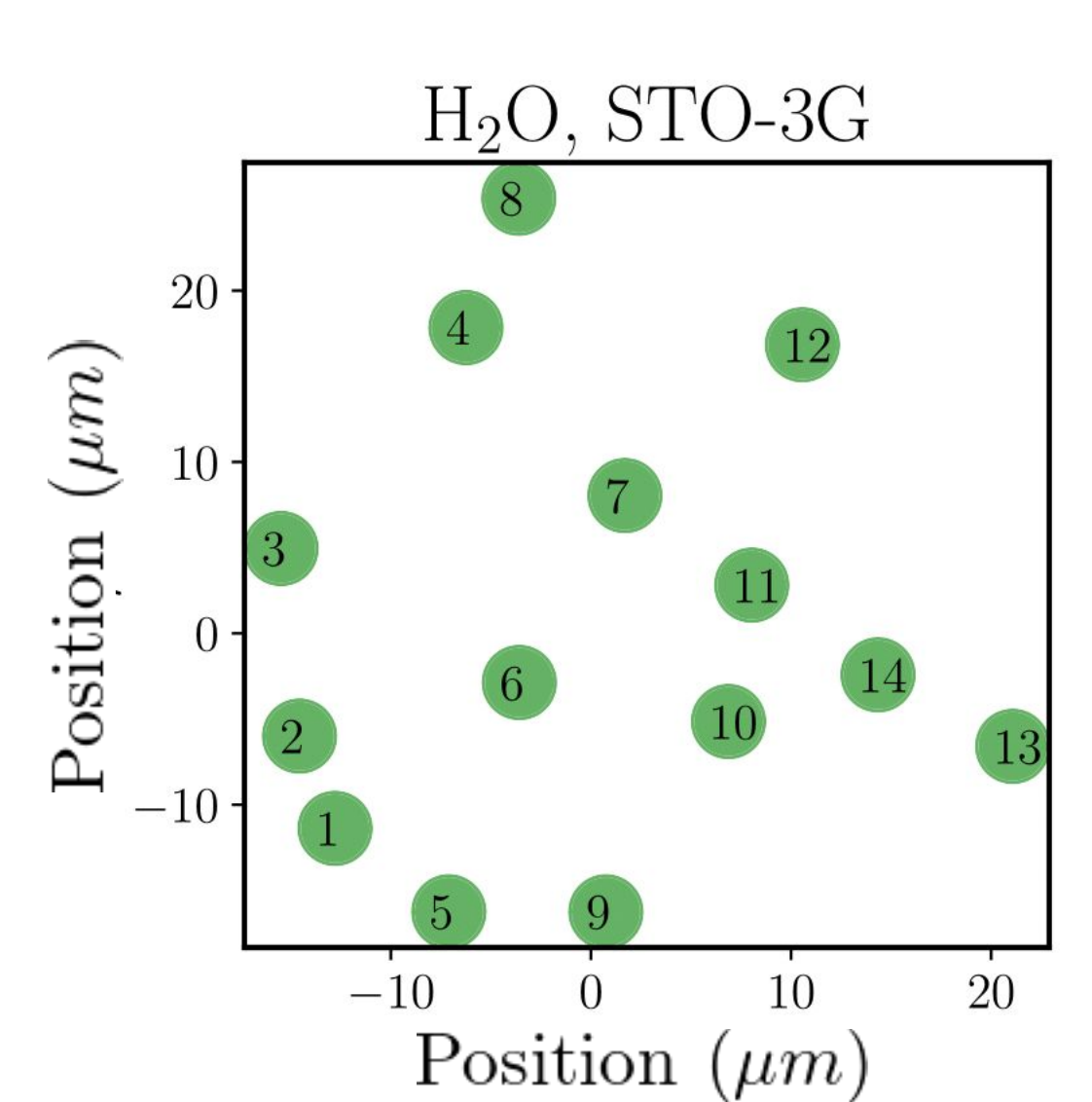}
    \includegraphics[width=0.49\linewidth,trim={5.5cm 2cm 8.5cm 1cm},clip]{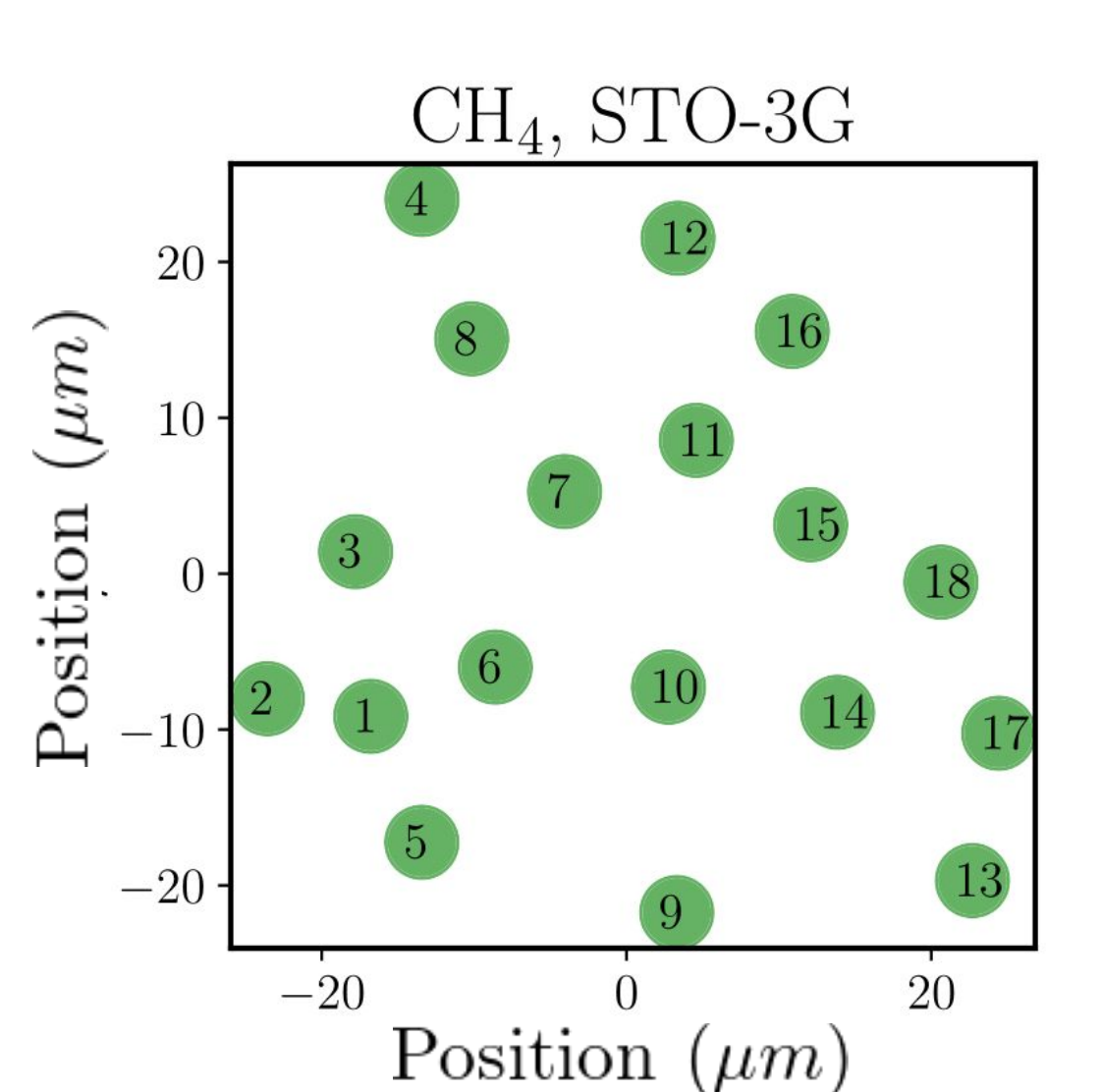}
    \includegraphics[width=0.49\linewidth,trim={5.5cm 1.8cm 8.5cm 1cm},clip]{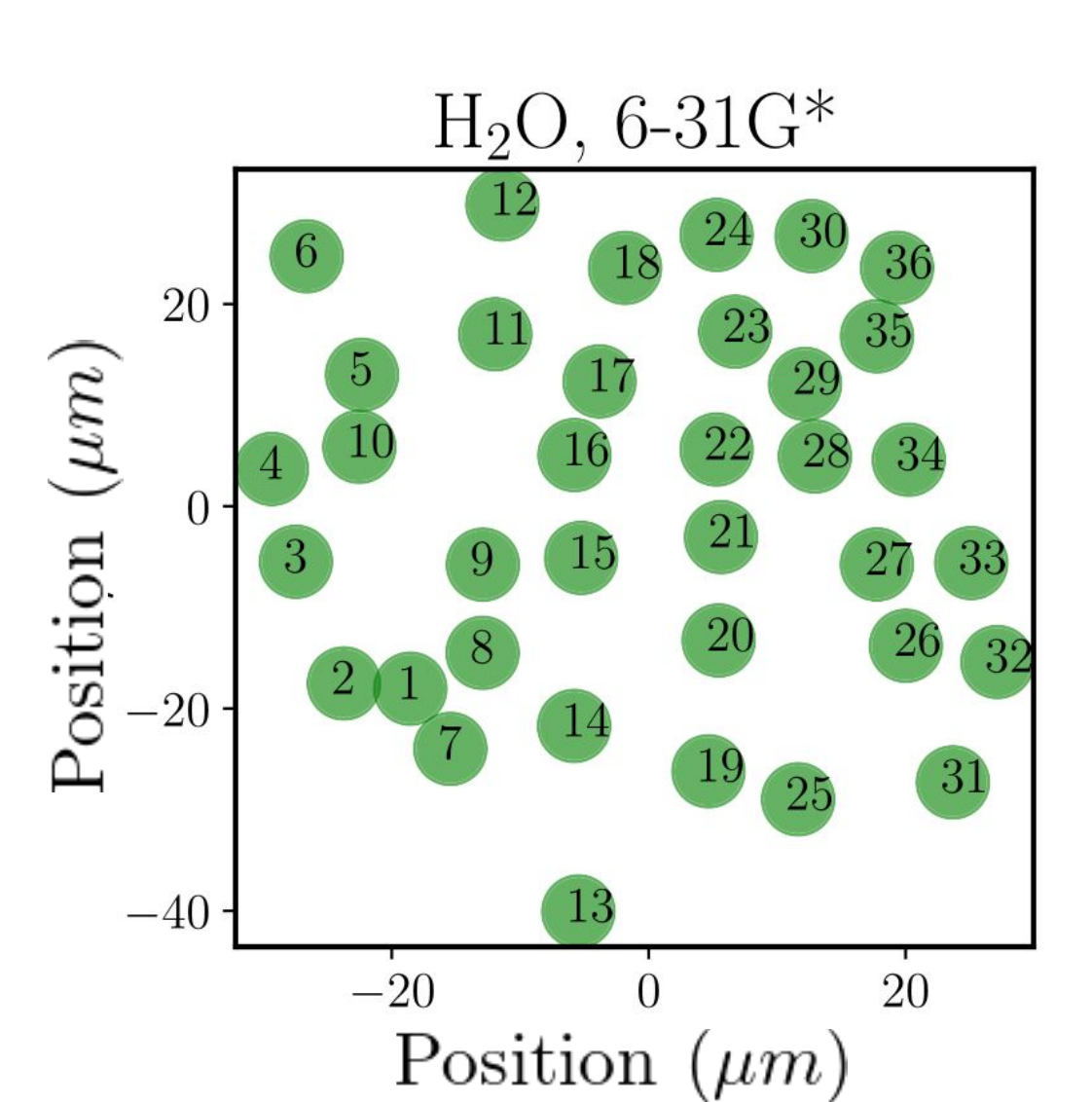}
    \includegraphics[width=0.48\linewidth,trim={5.5cm 2cm 8.5cm 1cm},clip]{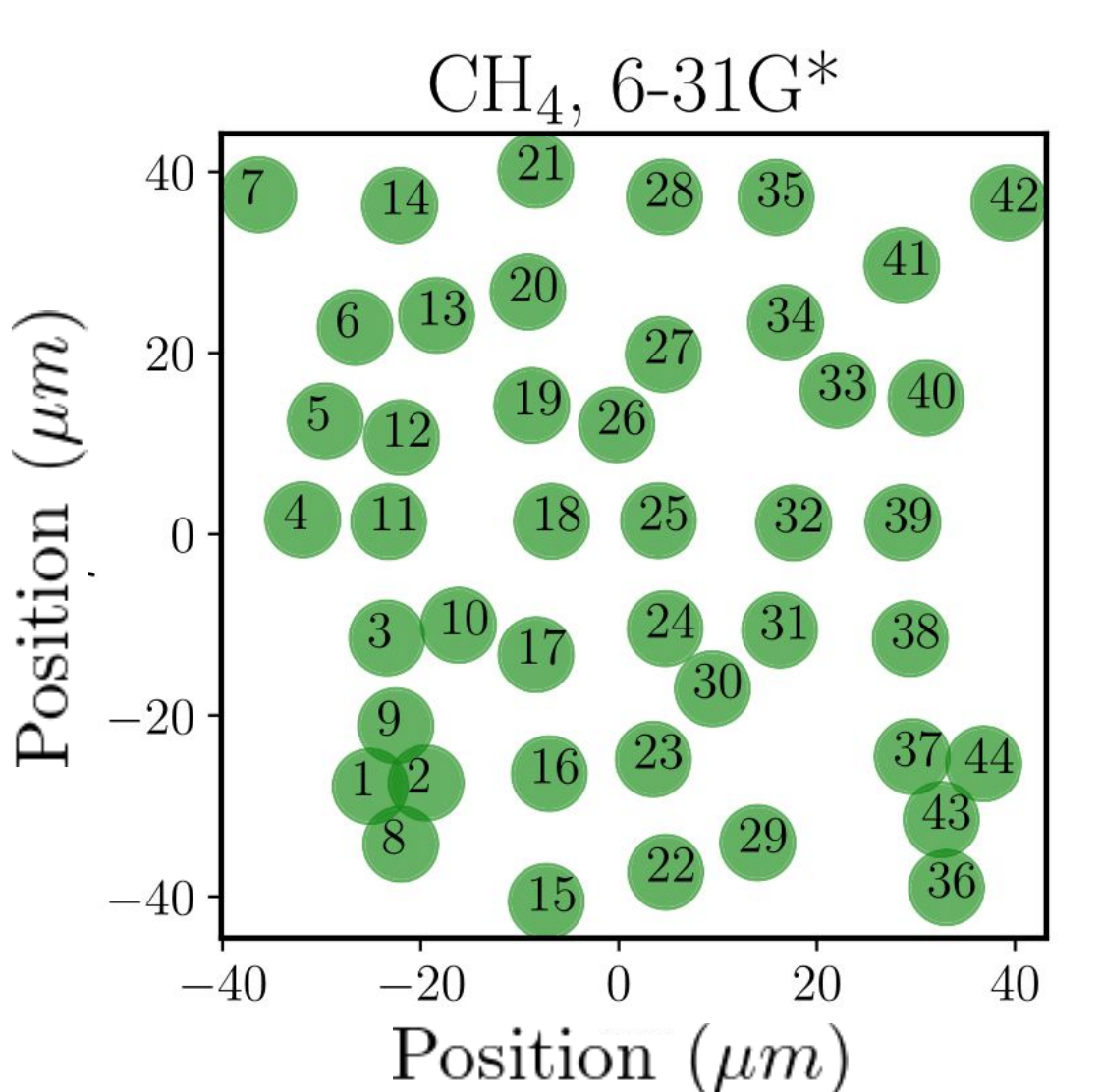}
    \caption{Example embeddings for the H$_2$O and CH$_4$ molecules. Using a Jordan-Wigner encoding, for two different basis, our H$_2$O Hamiltonian (left column) consisted of 14 qubits and 2110 terms (STO-3G basis), of which 595 were chosen as features for the reference Hamiltonian. Choosing a 6-31G* basis gives 36 qubits, with 83003 terms and 5594 of them relevant. For CH$_4$ (right column),  the STO-3G basis gives 18 atoms and 6892 Pauli terms, of which 1359 terms were used for the embedding. The 6-31G* basis requires 44 qubits, 297075 terms and we have picked an embedding with 11772 terms. }
    \label{fig:large_molecules}
\end{figure}

We discuss in this section some observations about the presented protocol for larger molecules and more complex basis sets, where the number of terms in the Hamiltonian and the required qubits to encode it grows quickly. Currently available neutral-atom devices can load hundreds of traps \cite{schymik2022situ}, but the available space on the register will eventually become a resource limitation. In Fig. \ref{fig:large_molecules} we show the embedding results for H$_2$O and CH$_4$ in different basis sets, where thousands of terms would need to be measured. Our simple restriction to $Z$-terms captures limited features of the Hamiltonian, mostly concentrating atoms where the largest values need to be reproduced. In fact, as the system size grows, we observe that most of the atoms in the register act as a ``background'' for these clusters. Choosing different terms from the Target Hamiltonian will bring forward other features, highlighting the opportunities of using learning methods to find more performing atom positions.  We have not addressed here the possibilities offered by  three-dimensional registers \cite{barredo2018}, which allow for more complex embeddings and have been already studied for graph-combinatorial problems \cite{dalyac2022embedding}, although they can be straightforwardly included in the protocol.

\begin{figure}[ht!]
\centering
\includegraphics[width=\linewidth,trim={0cm 0cm -0.5cm 0cm},clip]{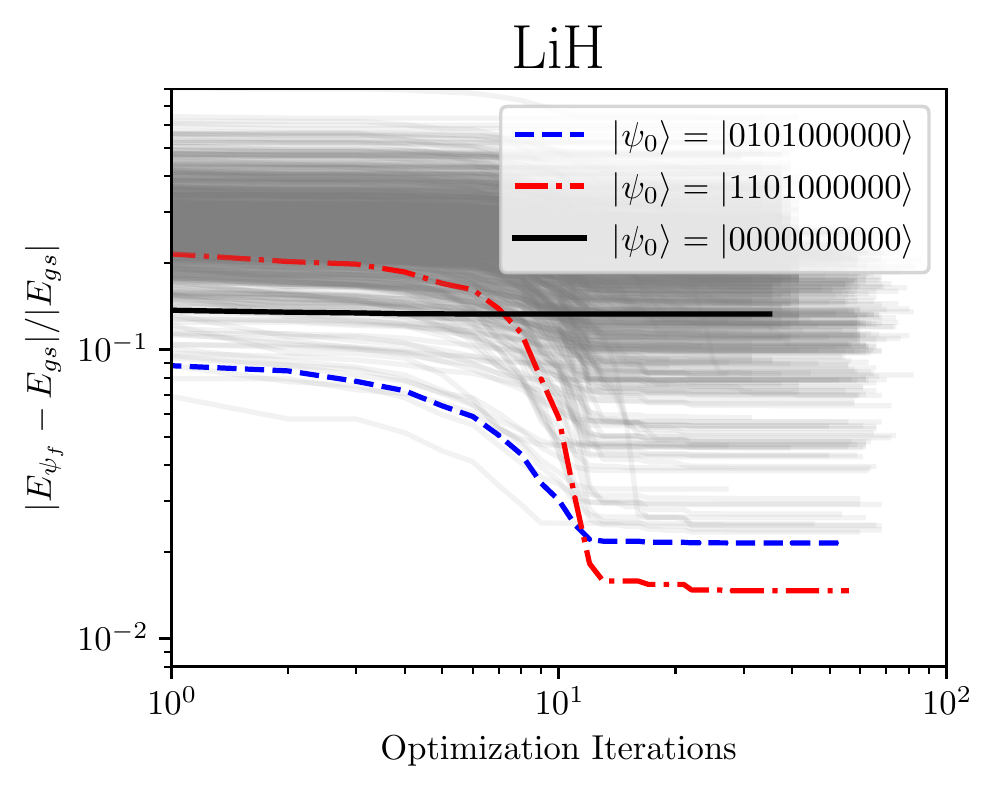}
\caption{Relative error $\varepsilon$ after the first step of the pulse optimization \ref{sec:pulse_opt_protocol}, using different initial product states. While the default $\ket{\psi_0}=\ket{0}^{\otimes N}$ (black line) barely improves its $\varepsilon$ by the optimization step, other product states (blue, dashed line and red, dashed-dotted line) achieve a very low error. We have used an 10-qubit Jordan-Wigner encoding of the LiH Molecular Hamiltonian under the STO-3G basis (276 terms), and averaged several instances for each product state.}
\label{fig:product_states}
\end{figure}

Two aspects of the optimization that rely on classical computation can be further refined: the choice of initial state and the selection of parameters: initializing the optimization with a product state from a Hartree-Fock state approximation can help exploring a lower energy set of output states. In Fig. \ref{fig:product_states} we have scanned through all product states of an 8-qubit system to select those who benefit the most from the first step of the pulse-optimization protocol presented above. The best choice of initial product state can then be used for the Rydberg QP implementation\footnote{Preparing the initial product state requires for example masking atoms with the help of a spatial light modulator.}. Comparing the energy $\langle \psi_0 | U(\theta)^\dagger \hat H_{\text{T}} U(\theta) | \psi_0 \rangle$ of a candidate initial product state $|\psi_0\rangle$ evolving under a constant pulse parameterized by $\theta$ can be performed for example using tensor-network techniques over an HPC backend \cite{bidzhiev2023emutn, rudolph_synergy_2022}. On the other hand, the number of parameters can be chosen at will, and do not depend on the number of qubits. More advanced control techniques can be applied here, and there is large choice of techniques and numerical tools in the subject of quantum optimal control. We recall that the optimization of the pulse sequence is not dependent on the embedding itself -- it is a global property of the system evolution, where  nearby atoms constitute blockade regions that characterize the final state. 

\section{Discussion}

In this work, we have numerically studied a digital-analog quantum algorithm in the context of quantum chemistry using an ideal Rydberg quantum processor as a hardware.  We have considered small molecules with resource Hamiltonians of two qubits for the H$_2$ molecule and six qubits for the LiH and BeH$_2$ molecules to demonstrate the applicability of our methods. Our purpose was to describe the construction of such an algorithm, discussing the cost  of each stage in terms of the number of measurement repetitions. Our numerical results should be viewed as a first benchmark and should trigger further explorations. Besides, it provides a roadmap for the improvement of Rydberg quantum processor, in particular in terms of cycle time. 

By considering the symmetries of H$_2$ Hamiltonian, we show how the UCC method efficiently and accurately approximates the ground state energy. However, finding a two-body Hamiltonian which commutes with a more general molecular target Hamiltonian is a hard problem. We therefore proposed another protocol for larger molecules: we optimized the geometries of the atomic array, pulse sequences and included an estimation method (derandomization) for the energy measurement. We targeted  $5\%$ of accuracy compared to the exact diagonalization method for Hamiltonian with 6 qubits and more than a hundred Pauli strings.

We observed that the geometry of the array has a significant impact on the result: In the case of LiH, where the target matrix $V^T$ does not provide much information due to the few terms with only two $Z$ operators, the optimized positions underperform with respect to a careful choice of positions, although scaling the heuristics that gives rise to such a geometry for molecules with large number of qubits in their encoding is impractical. This calls for the design of more advanced embedding algorithms. Indeed, for the case of BeH$_2$, the register optimization achieves rather small energy errors, especially for larger distances. 

Previous studies \cite{wecker_progress_2015} have quantified the demanding resource requirements for practical VQE applications. After several iterations of pulse optimization with energy estimation via derandomization, the error on the average energy of the final prepared state descends to $\varepsilon<5\%$, and expected to be obtained within a day of measurement in a typical current-day Rydberg QP.

In the numerical implementation we have used out-of-the-box optimizers with limitations for the numerical task at hand. Other possibilities include the use of an interpolated waveform for each set of parameters and shaping the pulse using bayesian optimization routines as  explored in \cite{coelho2022efficient} for the study of combinatorial graph problems. Note that experimentally, one could clone several times the atom layout in spatially separated regions of the register (at least for a small number of qubits), multiplying the obtained number of bitstrings. To achieve close to $1\%$ relative error, we expect that at least a week of Rydberg QP runtime would be necessary (see extrapolation shown in Fig. \ref{fig:results}). The capacity of a circuit ansatz to construct a desired quantum state while keeping a small depth and number of parameters is studied by its \emph{expressibility} \cite{sim2019expressibility, holmes2022connecting}. In the case of analog systems, this is an emerging topic of research \cite{tangpanitanon2020expressibility}, with the goal of ensuring that a given ansatz could potentially lead to a good approximation of the ground state and achieve chemical accuracy.

The impact of experimental errors in a real-life implementation will also lead to performance reductions. SPAM (State Preparation And Measurement) errors are typically the largest source of discrepancy for the neutral atom devices \cite{de_leseleuc_analysis_2018}, but the energy errors observed in numerical simulations remain low as long as the failure rates are small, given that the variational nature of the algorithm shows robustness to several types of errors \cite{henriet_robustness_2020}. Recently \cite{guo2022chemistry}, VQE was experimentally implemented in a superconducting quantum processor for H$_2$, LiH and F$_2$ with 4, 6 and 12 qubits, respectively, using the UCC ansatz and a different flavor of measurement protocol \cite{wu_overlapped_2023}. The readouts went through an error mitigation post-processing routine, which showed that these techniques can greatly compensate the noise effects from their quantum processor, with an error reduction of up to two orders of magnitude and leading to chemical accuracy in some circumstances. We expect that such mitigation can be added to the protocols considered in this paper and will help experimental implementations in Rydberg QP.

To tackle quantum simulation algorithms for energy estimation, more developments in both quantum and classical parts of the hybrid algorithm are needed. Reaching chemical accuracy for molecules with a few tens of qubits remains an open challenge that can now begin to be explored in experimental devices. This will provide evidence to generate new and fundamental insights to understand under what conditions a computational advantage can be achieved. 

\begin{acknowledgments}
We thank Thierry Lahaye and Louis Vignoli for discussions and reading of the manuscript.
This work was supported by the European Union's Horizon 2020 research and innovation 
program under grant  agreement No. 817482 (PASQuanS),  and the 
European Research Council (Advanced grant No. 101018511-ATARAXIA). It was also supported by EDF R\&D,
the Research and Development Division of Electricité de France under the ANRT contract N°2020/0011.
\end{acknowledgments}

\appendix
\section{LiH and BeH$_\text{2}$ Hamiltonians}
In this section, examples of complete Hamiltonians of molecules LiH (for an inter-atomic distance of $1.5 $ $\mathrm{\mathring{A}}$) and BeH$_\text{2}$ (for an inter-atomic distance of $1.17$ $ \mathrm{\mathring{A}}$) obtained with the method described in Sec.\ref{sec:chem} are shown.

\begin{widetext}\begin{equation}\begin{split}
H_\mathrm{LiH} =& {\bf -0.19975} +{\bf 0.05393}Z_0 {\bf -0.12836}Z_1 {\bf -0.31773}Z_0 Z_1 {\bf -0.31773}Z_3 +{\bf 0.0605}Z_1 Z_3  \\&+{\bf 0.11409}Z_0 Z_1 Z_3 +{\bf 0.05362}Z_4 +{\bf 0.11434}Z_2 Z_4 {\bf -0.03787}Z_2 Z_3 Z_4 +{\bf 0.05362}Z_1 Z_2 Z_3 Z_4  \\&+{\bf 0.0836}Z_0 Z_1 Z_2 Z_3 Z_4 {\bf -0.03787}Z_5 +{\bf 0.05666}Z_1 Z_5 +{\bf 0.11434}Z_0 Z_1 Z_5 +{\bf 0.0836}Z_3 Z_5  \\&+{\bf 0.05666}Z_2 Z_3 Z_5 {\bf -0.12836}Z_4 Z_5 +{\bf 0.0847}Z_1 Z_4 Z_5 +{\bf 0.0605}Z_0 Z_1 Z_4 Z_5 +{\bf 0.05393}Z_3 Z_4 Z_5  \\&+{\bf 0.12357}Z_2 Z_3 Z_4 Z_5 +{\bf 0.01522}X_1 {\bf -0.01522}Z_0 X_1 +{\bf 0.01089}X_1 Z_3 {\bf -0.01089}Z_0 X_1 Z_3  \\&+{\bf 0.00436}X_1 Z_2 Z_3 Z_4 {\bf -0.00436}Z_0 X_1 Z_2 Z_3 Z_4 +{\bf 0.01273}X_1 Z_5 {\bf -0.01273}Z_0 X_1 Z_5 {\bf -0.00901}X_1 Z_4 Z_5  \\&+{\bf 0.00901}Z_0 X_1 Z_4 Z_5 +{\bf 0.00448}X_0 X_2 {\bf -0.00479}X_0 Z_1 X_2 {\bf -0.03512}X_0 Z_1 X_2 Z_3 {\bf -0.03512}Y_0 Y_2 Z_4  \\&{\bf -0.00479}Y_0 Y_2 Z_3 Z_4 +{\bf 0.00448}Y_0 Z_1 Y_2 Z_3 Z_4 {\bf -0.03306}X_0 Z_1 X_2 Z_5 +{\bf 0.00237}Y_0 Y_2 Z_3 Z_5 +{\bf 0.00237}X_0 Z_1 X_2 Z_4 Z_5  \\&{\bf -0.03306}Y_0 Y_2 Z_3 Z_4 Z_5 - {\bf (4\times 10^{-5})}X_0 X_1 X_2 {\bf -0.00277}Y_0 Y_1 X_2 +{\bf 0.01054}X_0 X_1 X_2 Z_3 +{\bf 0.01054}X_0 Y_1 Y_2 Z_4  \\&+{\bf 0.00277}Y_0 X_1 Y_2 Z_3 Z_4 - {\bf (4\times 10^{-5})}X_0 Y_1 Y_2 Z_3 Z_4 +{\bf 0.01173}X_0 X_1 X_2 Z_5 {\bf -0.00154}X_0 Y_1 Y_2 Z_3 Z_5 \\&{\bf -0.00154}X_0 X_1 X_2 Z_4 Z_5  +{\bf 0.01173}X_0 Y_1 Y_2 Z_3 Z_4 Z_5 +{\bf 0.01522}X_3 X_4 {\bf -0.00901}Z_1 X_3 X_4 +{\bf 0.01089}Z_0 Z_1 X_3 X_4 \\& +{\bf 0.00436}Y_3 Y_4{\bf -0.01273}Z_2 Y_3 Y_4 +{\bf 0.00436}X_3 X_4 Z_5 {\bf -0.01273}Z_2 X_3 X_4 Z_5 +{\bf 0.01522}Y_3 Y_4 Z_5 {\bf -0.00901}Z_1 Y_3 Y_4 Z_5  \\&+{\bf 0.01089}Z_0 Z_1 Y_3 Y_4 Z_5 +{\bf 0.00658}X_1 X_3 X_4 {\bf -0.00658}Z_0 X_1 X_3 X_4 +{\bf 0.00658}X_1 Y_3 Y_4 Z_5 {\bf -0.00658}Z_0 X_1 Y_3 Y_4 Z_5  \\&{\bf -0.00776}X_0 Z_1 X_2 X_3 X_4 +{\bf 0.00776}Y_0 Y_2 Y_3 Y_4 +{\bf 0.00776}Y_0 Y_2 X_3 X_4 Z_5 {\bf -0.00776}X_0 Z_1 X_2 Y_3 Y_4 Z_5 \\&+{\bf 0.00211}X_0 X_1 X_2 X_3 X_4  {\bf -0.00211}X_0 Y_1 Y_2 Y_3 Y_4 {\bf -0.00211}X_0 Y_1 Y_2 X_3 X_4 Z_5 +{\bf 0.00211}X_0 X_1 X_2 Y_3 Y_4 Z_5 + {\bf 0.00004}X_5 \\&{\bf -0.00154}Z_1 X_5  +{\bf 0.01054}Z_0 Z_1 X_5 +{\bf 0.00277}Z_3 X_5 {\bf -0.01173}Z_2 Z_3 X_5 +{\bf 0.00004}Z_4 X_5 +{\bf 0.00154}Z_1 Z_4 X_5  \\&{\bf -0.01054}Z_0 Z_1 Z_4 X_5 {\bf -0.00277}Z_3 Z_4 X_5 +{\bf 0.01173}Z_2 Z_3 Z_4 X_5 +{\bf 0.00211}X_1 X_5 {\bf -0.00211}Z_0 X_1 X_5  \\&{\bf -0.00211}X_1 Z_4 X_5 +{\bf 0.00211}Z_0 X_1 Z_4 X_5 {\bf -0.00837}X_0 Z_1 X_2 X_5 +{\bf 0.00837}Y_0 Y_2 Z_3 X_5 +{\bf 0.00837}X_0 Z_1 X_2 Z_4 X_5  \\&{\bf -0.00837}Y_0 Y_2 Z_3 Z_4 X_5 +{\bf 0.00303}X_0 X_1 X_2 X_5 {\bf -0.00303}X_0 Y_1 Y_2 Z_3 X_5 {\bf -0.00303}X_0 X_1 X_2 Z_4 X_5 \\&+{\bf 0.00303}X_0 Y_1 Y_2 Z_3 Z_4 X_5  +{\bf 0.00448}X_3 X_4 X_5 +{\bf 0.03306}Z_2 X_3 X_4 X_5 {\bf -0.00479}Y_3 Y_4 X_5 +{\bf 0.00237}Z_1 Y_3 Y_4 X_5 \\& {\bf -0.03512}Z_0 Z_1 Y_3 Y_4 X_5  +{\bf 0.00448}Y_3 X_4 Y_5 +{\bf 0.03306}Z_2 Y_3 X_4 Y_5 +{\bf 0.00479}X_3 Y_4 Y_5 {\bf -0.00237}Z_1 X_3 Y_4 Y_5 \\& +{\bf 0.03512}Z_0 Z_1 X_3 Y_4 Y_5  {\bf -0.00776}X_1 Y_3 Y_4 X_5 +{\bf 0.00776}Z_0 X_1 Y_3 Y_4 X_5 +{\bf 0.00776}X_1 X_3 Y_4 Y_5 {\bf -0.00776}Z_0 X_1 X_3 Y_4 Y_5 \\&{\bf -0.03074}Y_0 Y_2 X_3 X_4 X_5  +{\bf 0.03074}X_0 Z_1 X_2 Y_3 Y_4 X_5 {\bf -0.03074}Y_0 Y_2 Y_3 X_4 Y_5 {\bf -0.03074}X_0 Z_1 X_2 X_3 Y_4 Y_5 \\&+{\bf 0.00837}X_0 Y_1 Y_2 X_3 X_4 X_5  {\bf -0.00837}X_0 X_1 X_2 Y_3 Y_4 X_5  +{\bf 0.00837}X_0 Y_1 Y_2 Y_3 X_4 Y_5 +{\bf 0.00837}X_0 X_1 X_2 X_3 Y_4 Y_5   
\end{split}
\end{equation}
\end{widetext}
\begin{widetext}
\begin{equation}
\begin{split}
H_{\text{BeH}_\text{2}} =& {\bf -1.90305} {\bf -0.48894}Z_0 +{\bf 0.14357}Z_1 {\bf -0.18803}Z_0 Z_1 +{\bf 0.12314}Z_2 +{\bf 0.18326}Z_0 Z_2  \\&+{\bf 0.10964}Z_1 Z_2 +{\bf 0.18222}Z_0 Z_1 Z_2 {\bf -0.48894}Z_3 +{\bf 0.1288}Z_0 Z_3 +{\bf 0.1136}Z_0 Z_1 Z_3  \\&+{\bf 0.11249}Z_2 Z_3 +{\bf 0.11746}Z_1 Z_2 Z_3 +{\bf 0.14357}Z_4 {\bf -0.18803}Z_3 Z_4 +{\bf 0.1136}Z_0 Z_3 Z_4  \\&+{\bf 0.10602}Z_0 Z_1 Z_3 Z_4 +{\bf 0.10306}Z_2 Z_3 Z_4 +{\bf 0.10577}Z_1 Z_2 Z_3 Z_4 +{\bf 0.12314}Z_5 +{\bf 0.11249}Z_0 Z_5  \\&+{\bf 0.10306}Z_0 Z_1 Z_5 +{\bf 0.10451}Z_2 Z_5 +{\bf 0.10785}Z_1 Z_2 Z_5 +{\bf 0.18326}Z_3 Z_5 +{\bf 0.10964}Z_4 Z_5  \\&+{\bf 0.11746}Z_0 Z_4 Z_5 +{\bf 0.10577}Z_0 Z_1 Z_4 Z_5 +{\bf 0.10785}Z_2 Z_4 Z_5 +{\bf 0.11352}Z_1 Z_2 Z_4 Z_5 +{\bf 0.18222}Z_3 Z_4 Z_5  \\&{\bf -0.00743}X_0 X_1 {\bf -0.00229}Y_0 Y_1 {\bf -0.00229}X_0 X_1 Z_2 {\bf -0.00743}Y_0 Y_1 Z_2 {\bf -0.00711}Y_0 Y_1 Z_3  \\&{\bf -0.00711}X_0 X_1 Z_2 Z_3 {\bf -0.00875}Y_0 Y_1 Z_3 Z_4 {\bf -0.00875}X_0 X_1 Z_2 Z_3 Z_4 {\bf -0.00352}Y_0 Y_1 Z_5 {\bf -0.00352}X_0 X_1 Z_2 Z_5  \\&{\bf -0.0072}Y_0 Y_1 Z_4 Z_5 {\bf -0.0072}X_0 X_1 Z_2 Z_4 Z_5 {\bf -0.04165}X_0 X_2 +{\bf 0.03769}X_0 Z_1 X_2 {\bf -0.00396}Y_0 Y_2  \\&+{\bf 0.00839}X_1 X_2 {\bf -0.01015}Z_0 X_1 X_2 {\bf -0.01015}Y_1 Y_2 +{\bf 0.00839}Z_0 Y_1 Y_2 +{\bf 0.01355}Z_0 X_1 X_2 Z_3  \\&+{\bf 0.01355}Y_1 Y_2 Z_3 +{\bf 0.01082}Z_0 X_1 X_2 Z_3 Z_4 +{\bf 0.01082}Y_1 Y_2 Z_3 Z_4 +{\bf 0.00854}Z_0 X_1 X_2 Z_5 +{\bf 0.00854}Y_1 Y_2 Z_5  \\&+{\bf 0.01408}Z_0 X_1 X_2 Z_4 Z_5 +{\bf 0.01408}Y_1 Y_2 Z_4 Z_5 +{\bf 0.03611}X_0 X_3 {\bf -0.03611}X_0 Z_1 X_3 {\bf -0.03611}X_0 X_3 Z_4  \\&+{\bf 0.03611}X_0 Z_1 X_3 Z_4 {\bf -0.02498}X_1 X_3 +{\bf 0.02498}Z_0 X_1 Z_2 X_3 +{\bf 0.02498}X_1 X_3 Z_4 {\bf -0.02498}Z_0 X_1 Z_2 X_3 Z_4  \\&{\bf -0.03615}X_2 X_3 +{\bf 0.03615}Z_1 X_2 X_3 +{\bf 0.03615}X_2 X_3 Z_4 {\bf -0.03615}Z_1 X_2 X_3 Z_4 {\bf -0.01573}X_0 X_1 X_2 X_3  \\&{\bf -0.01573}Y_0 X_1 Y_2 X_3 +{\bf 0.01573}X_0 X_1 X_2 X_3 Z_4 +{\bf 0.01573}Y_0 X_1 Y_2 X_3 Z_4 {\bf -0.02498}X_0 X_4 +{\bf 0.02498}X_0 Z_1 X_4  \\&+{\bf 0.02498}X_0 Z_3 X_4 Z_5 {\bf -0.02498}X_0 Z_1 Z_3 X_4 Z_5 +{\bf 0.02085}X_1 X_4 {\bf -0.02085}Z_0 X_1 Z_2 X_4 {\bf -0.02085}X_1 Z_3 X_4 Z_5  \\&+{\bf 0.02085}Z_0 X_1 Z_2 Z_3 X_4 Z_5 +{\bf 0.02464}X_2 X_4 {\bf -0.02464}Z_1 X_2 X_4 {\bf -0.02464}X_2 Z_3 X_4 Z_5 +{\bf 0.02464}Z_1 X_2 Z_3 X_4 Z_5  \\&+{\bf 0.01532}X_0 X_1 X_2 X_4 +{\bf 0.01532}Y_0 X_1 Y_2 X_4 {\bf -0.01532}X_0 X_1 X_2 Z_3 X_4 Z_5 {\bf -0.01532}Y_0 X_1 Y_2 Z_3 X_4 Z_5 {\bf -0.00743}X_3 X_4  \\&{\bf -0.00229}Y_3 Y_4 {\bf -0.00711}Z_0 Y_3 Y_4 {\bf -0.00875}Z_0 Z_1 Y_3 Y_4 {\bf -0.00352}Z_2 Y_3 Y_4 {\bf -0.0072}Z_1 Z_2 Y_3 Y_4  \\&{\bf -0.00229}X_3 X_4 Z_5 {\bf -0.00711}Z_0 X_3 X_4 Z_5 {\bf -0.00875}Z_0 Z_1 X_3 X_4 Z_5 {\bf -0.00352}Z_2 X_3 X_4 Z_5 {\bf -0.0072}Z_1 Z_2 X_3 X_4 Z_5  \\&{\bf -0.00743}Y_3 Y_4 Z_5 +{\bf 0.01972}Y_0 Y_1 Y_3 Y_4 +{\bf 0.01972}X_0 X_1 Z_2 Y_3 Y_4 +{\bf 0.01972}Y_0 Y_1 X_3 X_4 Z_5 +{\bf 0.01972}X_0 X_1 Z_2 X_3 X_4 Z_5  \\&{\bf -0.0173}Z_0 X_1 X_2 Y_3 Y_4 {\bf -0.0173}Y_1 Y_2 Y_3 Y_4 {\bf -0.0173}Z_0 X_1 X_2 X_3 X_4 Z_5 {\bf -0.0173}Y_1 Y_2 X_3 X_4 Z_5 {\bf -0.03615}X_0 X_5  \\&+{\bf 0.03615}X_0 Z_1 X_5 +{\bf 0.03615}X_0 Z_4 X_5 {\bf -0.03615}X_0 Z_1 Z_4 X_5 +{\bf 0.02464}X_1 X_5 {\bf -0.02464}Z_0 X_1 Z_2 X_5  \\&{\bf -0.02464}X_1 Z_4 X_5 +{\bf 0.02464}Z_0 X_1 Z_2 Z_4 X_5 +{\bf 0.04177}X_2 X_5 {\bf -0.04177}Z_1 X_2 X_5 {\bf -0.04177}X_2 Z_4 X_5  \\&+{\bf 0.04177}Z_1 X_2 Z_4 X_5 +{\bf 0.01232}X_0 X_1 X_2 X_5 +{\bf 0.01232}Y_0 X_1 Y_2 X_5 {\bf -0.01232}X_0 X_1 X_2 Z_4 X_5 {\bf -0.01232}Y_0 X_1 Y_2 Z_4 X_5  \\&{\bf -0.04165}X_3 X_5 +{\bf 0.03769}X_3 Z_4 X_5 {\bf -0.00396}Y_3 Y_5 +{\bf 0.00839}X_4 X_5 {\bf -0.01015}Z_3 X_4 X_5  \\&+{\bf 0.01355}Z_0 Z_3 X_4 X_5 +{\bf 0.01082}Z_0 Z_1 Z_3 X_4 X_5 +{\bf 0.00854}Z_2 Z_3 X_4 X_5 +{\bf 0.01408}Z_1 Z_2 Z_3 X_4 X_5 {\bf -0.01015}Y_4 Y_5  \\&+{\bf 0.01355}Z_0 Y_4 Y_5 +{\bf 0.01082}Z_0 Z_1 Y_4 Y_5 +{\bf 0.00854}Z_2 Y_4 Y_5 +{\bf 0.01408}Z_1 Z_2 Y_4 Y_5 +{\bf 0.00839}Z_3 Y_4 Y_5  \\&{\bf -0.0173}Y_0 Y_1 Z_3 X_4 X_5 {\bf -0.0173}X_0 X_1 Z_2 Z_3 X_4 X_5 {\bf -0.0173}Y_0 Y_1 Y_4 Y_5 {\bf -0.0173}X_0 X_1 Z_2 Y_4 Y_5 +{\bf 0.01858}Z_0 X_1 X_2 Z_3 X_4 X_5  \\&+{\bf 0.01858}Y_1 Y_2 Z_3 X_4 X_5 +{\bf 0.01858}Z_0 X_1 X_2 Y_4 Y_5 +{\bf 0.01858}Y_1 Y_2 Y_4 Y_5 {\bf -0.01573}X_0 X_3 X_4 X_5 \\&+{\bf 0.01573}X_0 Z_1 X_3 X_4 X_5  {\bf -0.01573}X_0 Y_3 X_4 Y_5 +{\bf 0.01573}X_0 Z_1 Y_3 X_4 Y_5 +{\bf 0.01532}X_1 X_3 X_4 X_5 {\bf -0.01532}Z_0 X_1 Z_2 X_3 X_4 X_5 +{\bf 0.01532}X_1 Y_3 X_4 Y_5 \\& {\bf -0.01532}Z_0 X_1 Z_2 Y_3 X_4 Y_5   +{\bf 0.01232}X_2 X_3 X_4 X_5 {\bf -0.01232}Z_1 X_2 X_3 X_4 X_5 +{\bf 0.01232}X_2 Y_3 X_4 Y_5 \\& {\bf -0.01232}Z_1 X_2 Y_3 X_4 Y_5  +{\bf 0.01415}X_0 X_1 X_2 X_3 X_4 X_5 +{\bf 0.01415}Y_0 X_1 Y_2 X_3 X_4 X_5 +{\bf 0.01415}X_0 X_1 X_2 Y_3 X_4 Y_5 \\& +{\bf 0.01415}Y_0 X_1 Y_2 Y_3 X_4 Y_5 
\end{split}
\end{equation}
\end{widetext}

\bibliography{main.bib}

\begin{thebibliography}{56}%
\makeatletter
\providecommand \@ifxundefined [1]{%
 \@ifx{#1\undefined}
}%
\providecommand \@ifnum [1]{%
 \ifnum #1\expandafter \@firstoftwo
 \else \expandafter \@secondoftwo
 \fi
}%
\providecommand \@ifx [1]{%
 \ifx #1\expandafter \@firstoftwo
 \else \expandafter \@secondoftwo
 \fi
}%
\providecommand \natexlab [1]{#1}%
\providecommand \enquote  [1]{``#1''}%
\providecommand \bibnamefont  [1]{#1}%
\providecommand \bibfnamefont [1]{#1}%
\providecommand \citenamefont [1]{#1}%
\providecommand \href@noop [0]{\@secondoftwo}%
\providecommand \href [0]{\begingroup \@sanitize@url \@href}%
\providecommand \@href[1]{\@@startlink{#1}\@@href}%
\providecommand \@@href[1]{\endgroup#1\@@endlink}%
\providecommand \@sanitize@url [0]{\catcode `\\12\catcode `\$12\catcode
  `\&12\catcode `\#12\catcode `\^12\catcode `\_12\catcode `\%12\relax}%
\providecommand \@@startlink[1]{}%
\providecommand \@@endlink[0]{}%
\providecommand \url  [0]{\begingroup\@sanitize@url \@url }%
\providecommand \@url [1]{\endgroup\@href {#1}{\urlprefix }}%
\providecommand \urlprefix  [0]{URL }%
\providecommand \Eprint [0]{\href }%
\providecommand \doibase [0]{https://doi.org/}%
\providecommand \selectlanguage [0]{\@gobble}%
\providecommand \bibinfo  [0]{\@secondoftwo}%
\providecommand \bibfield  [0]{\@secondoftwo}%
\providecommand \translation [1]{[#1]}%
\providecommand \BibitemOpen [0]{}%
\providecommand \bibitemStop [0]{}%
\providecommand \bibitemNoStop [0]{.\EOS\space}%
\providecommand \EOS [0]{\spacefactor3000\relax}%
\providecommand \BibitemShut  [1]{\csname bibitem#1\endcsname}%
\let\auto@bib@innerbib\@empty
\bibitem [{\citenamefont {Georgescu}\ \emph {et~al.}(2014)\citenamefont
  {Georgescu}, \citenamefont {Ashhab},\ and\ \citenamefont
  {Nori}}]{georgescu_quantum_2014}%
  \BibitemOpen
  \bibfield  {author} {\bibinfo {author} {\bibfnamefont {I.}~\bibnamefont
  {Georgescu}}, \bibinfo {author} {\bibfnamefont {S.}~\bibnamefont {Ashhab}},\
  and\ \bibinfo {author} {\bibfnamefont {F.}~\bibnamefont {Nori}},\ }\bibfield
  {title} {\bibinfo {title} {Quantum simulation},\ }\href
  {https://doi.org/10.1103/RevModPhys.86.153} {\bibfield  {journal} {\bibinfo
  {journal} {Reviews of Modern Physics}\ }\textbf {\bibinfo {volume} {86}},\
  \bibinfo {pages} {153} (\bibinfo {year} {2014})}\BibitemShut {NoStop}%
\bibitem [{\citenamefont {McClean}\ \emph {et~al.}(2016)\citenamefont
  {McClean}, \citenamefont {Romero}, \citenamefont {Babbush},\ and\
  \citenamefont {Aspuru-Guzik}}]{mcclean_theory_2016}%
  \BibitemOpen
  \bibfield  {author} {\bibinfo {author} {\bibfnamefont {J.~R.}\ \bibnamefont
  {McClean}}, \bibinfo {author} {\bibfnamefont {J.}~\bibnamefont {Romero}},
  \bibinfo {author} {\bibfnamefont {R.}~\bibnamefont {Babbush}},\ and\ \bibinfo
  {author} {\bibfnamefont {A.}~\bibnamefont {Aspuru-Guzik}},\ }\bibfield
  {title} {\bibinfo {title} {The theory of variational hybrid quantum-classical
  algorithms},\ }\href {https://doi.org/10.1088/1367-2630/18/2/023023}
  {\bibfield  {journal} {\bibinfo  {journal} {New Journal of Physics}\ }\textbf
  {\bibinfo {volume} {18}},\ \bibinfo {pages} {023023} (\bibinfo {year}
  {2016})}\BibitemShut {NoStop}%
\bibitem [{\citenamefont {Lanyon}\ \emph {et~al.}(2010)\citenamefont {Lanyon},
  \citenamefont {Whitfield}, \citenamefont {Gillett}, \citenamefont {Goggin},
  \citenamefont {Almeida}, \citenamefont {Kassal}, \citenamefont {Biamonte},
  \citenamefont {Mohseni}, \citenamefont {Powell}, \citenamefont {Barbieri},
  \citenamefont {Aspuru-Guzik},\ and\ \citenamefont {White}}]{Lanyon2010}%
  \BibitemOpen
  \bibfield  {author} {\bibinfo {author} {\bibfnamefont {B.~P.}\ \bibnamefont
  {Lanyon}}, \bibinfo {author} {\bibfnamefont {J.~D.}\ \bibnamefont
  {Whitfield}}, \bibinfo {author} {\bibfnamefont {G.~G.}\ \bibnamefont
  {Gillett}}, \bibinfo {author} {\bibfnamefont {M.~E.}\ \bibnamefont {Goggin}},
  \bibinfo {author} {\bibfnamefont {M.~P.}\ \bibnamefont {Almeida}}, \bibinfo
  {author} {\bibfnamefont {I.}~\bibnamefont {Kassal}}, \bibinfo {author}
  {\bibfnamefont {J.~D.}\ \bibnamefont {Biamonte}}, \bibinfo {author}
  {\bibfnamefont {M.}~\bibnamefont {Mohseni}}, \bibinfo {author} {\bibfnamefont
  {B.~J.}\ \bibnamefont {Powell}}, \bibinfo {author} {\bibfnamefont
  {M.}~\bibnamefont {Barbieri}}, \bibinfo {author} {\bibfnamefont
  {A.}~\bibnamefont {Aspuru-Guzik}},\ and\ \bibinfo {author} {\bibfnamefont
  {A.~G.}\ \bibnamefont {White}},\ }\bibfield  {title} {\bibinfo {title}
  {Towards quantum chemistry on a quantum computer},\ }\href
  {https://doi.org/10.1038/nchem.483} {\bibfield  {journal} {\bibinfo
  {journal} {Nature Chemistry}\ }\textbf {\bibinfo {volume} {2}},\ \bibinfo
  {pages} {106} (\bibinfo {year} {2010})}\BibitemShut {NoStop}%
\bibitem [{\citenamefont {Peruzzo}\ \emph {et~al.}(2014)\citenamefont
  {Peruzzo}, \citenamefont {McClean}, \citenamefont {Shadbolt}, \citenamefont
  {Yung}, \citenamefont {Zhou}, \citenamefont {Love}, \citenamefont
  {Aspuru-Guzik},\ and\ \citenamefont {O'Brien}}]{Peruzzo2014}%
  \BibitemOpen
  \bibfield  {author} {\bibinfo {author} {\bibfnamefont {A.}~\bibnamefont
  {Peruzzo}}, \bibinfo {author} {\bibfnamefont {J.}~\bibnamefont {McClean}},
  \bibinfo {author} {\bibfnamefont {P.}~\bibnamefont {Shadbolt}}, \bibinfo
  {author} {\bibfnamefont {M.-H.}\ \bibnamefont {Yung}}, \bibinfo {author}
  {\bibfnamefont {X.-Q.}\ \bibnamefont {Zhou}}, \bibinfo {author}
  {\bibfnamefont {P.~J.}\ \bibnamefont {Love}}, \bibinfo {author}
  {\bibfnamefont {A.}~\bibnamefont {Aspuru-Guzik}},\ and\ \bibinfo {author}
  {\bibfnamefont {J.~L.}\ \bibnamefont {O'Brien}},\ }\bibfield  {title}
  {\bibinfo {title} {A variational eigenvalue solver on a photonic quantum
  processor},\ }\href {https://doi.org/10.1038/ncomms5213} {\bibfield
  {journal} {\bibinfo  {journal} {Nature Communications}\ }\textbf {\bibinfo
  {volume} {5}},\ \bibinfo {pages} {4213} (\bibinfo {year} {2014})}\BibitemShut
  {NoStop}%
\bibitem [{\citenamefont {Shen}\ \emph {et~al.}(2017)\citenamefont {Shen},
  \citenamefont {Zhang}, \citenamefont {Zhang}, \citenamefont {Zhang},
  \citenamefont {Yung},\ and\ \citenamefont {Kim}}]{Shen2017}%
  \BibitemOpen
  \bibfield  {author} {\bibinfo {author} {\bibfnamefont {Y.}~\bibnamefont
  {Shen}}, \bibinfo {author} {\bibfnamefont {X.}~\bibnamefont {Zhang}},
  \bibinfo {author} {\bibfnamefont {S.}~\bibnamefont {Zhang}}, \bibinfo
  {author} {\bibfnamefont {J.-N.}\ \bibnamefont {Zhang}}, \bibinfo {author}
  {\bibfnamefont {M.-H.}\ \bibnamefont {Yung}},\ and\ \bibinfo {author}
  {\bibfnamefont {K.}~\bibnamefont {Kim}},\ }\bibfield  {title} {\bibinfo
  {title} {Quantum implementation of the unitary coupled cluster for simulating
  molecular electronic structure},\ }\href
  {https://doi.org/10.1103/PhysRevA.95.020501} {\bibfield  {journal} {\bibinfo
  {journal} {Phys. Rev. A}\ }\textbf {\bibinfo {volume} {95}},\ \bibinfo
  {pages} {020501} (\bibinfo {year} {2017})}\BibitemShut {NoStop}%
\bibitem [{\citenamefont {Shen}\ \emph {et~al.}(2018)\citenamefont {Shen},
  \citenamefont {Lu}, \citenamefont {Zhang}, \citenamefont {Zhang},
  \citenamefont {Zhang}, \citenamefont {Huh},\ and\ \citenamefont
  {Kim}}]{Shen2018}%
  \BibitemOpen
  \bibfield  {author} {\bibinfo {author} {\bibfnamefont {Y.}~\bibnamefont
  {Shen}}, \bibinfo {author} {\bibfnamefont {Y.}~\bibnamefont {Lu}}, \bibinfo
  {author} {\bibfnamefont {K.}~\bibnamefont {Zhang}}, \bibinfo {author}
  {\bibfnamefont {J.}~\bibnamefont {Zhang}}, \bibinfo {author} {\bibfnamefont
  {S.}~\bibnamefont {Zhang}}, \bibinfo {author} {\bibfnamefont
  {J.}~\bibnamefont {Huh}},\ and\ \bibinfo {author} {\bibfnamefont
  {K.}~\bibnamefont {Kim}},\ }\bibfield  {title} {\bibinfo {title} {Quantum
  optical emulation of molecular vibronic spectroscopy using a trapped-ion
  device},\ }\href {https://doi.org/10.1039/C7SC04602B} {\bibfield  {journal}
  {\bibinfo  {journal} {Chem. Sci.}\ }\textbf {\bibinfo {volume} {9}},\
  \bibinfo {pages} {836} (\bibinfo {year} {2018})}\BibitemShut {NoStop}%
\bibitem [{\citenamefont {Hempel}\ \emph {et~al.}(2018)\citenamefont {Hempel},
  \citenamefont {Maier}, \citenamefont {Romero}, \citenamefont {McClean},
  \citenamefont {Monz}, \citenamefont {Shen}, \citenamefont {Jurcevic},
  \citenamefont {Lanyon}, \citenamefont {Love}, \citenamefont {Babbush},
  \citenamefont {Aspuru-Guzik}, \citenamefont {Blatt},\ and\ \citenamefont
  {Roos}}]{Hempel2018}%
  \BibitemOpen
  \bibfield  {author} {\bibinfo {author} {\bibfnamefont {C.}~\bibnamefont
  {Hempel}}, \bibinfo {author} {\bibfnamefont {C.}~\bibnamefont {Maier}},
  \bibinfo {author} {\bibfnamefont {J.}~\bibnamefont {Romero}}, \bibinfo
  {author} {\bibfnamefont {J.}~\bibnamefont {McClean}}, \bibinfo {author}
  {\bibfnamefont {T.}~\bibnamefont {Monz}}, \bibinfo {author} {\bibfnamefont
  {H.}~\bibnamefont {Shen}}, \bibinfo {author} {\bibfnamefont {P.}~\bibnamefont
  {Jurcevic}}, \bibinfo {author} {\bibfnamefont {B.~P.}\ \bibnamefont
  {Lanyon}}, \bibinfo {author} {\bibfnamefont {P.}~\bibnamefont {Love}},
  \bibinfo {author} {\bibfnamefont {R.}~\bibnamefont {Babbush}}, \bibinfo
  {author} {\bibfnamefont {A.}~\bibnamefont {Aspuru-Guzik}}, \bibinfo {author}
  {\bibfnamefont {R.}~\bibnamefont {Blatt}},\ and\ \bibinfo {author}
  {\bibfnamefont {C.~F.}\ \bibnamefont {Roos}},\ }\bibfield  {title} {\bibinfo
  {title} {Quantum chemistry calculations on a trapped-ion quantum simulator},\
  }\href {https://doi.org/10.1103/PhysRevX.8.031022} {\bibfield  {journal}
  {\bibinfo  {journal} {Phys. Rev. X}\ }\textbf {\bibinfo {volume} {8}},\
  \bibinfo {pages} {031022} (\bibinfo {year} {2018})}\BibitemShut {NoStop}%
\bibitem [{\citenamefont {Kandala}\ \emph {et~al.}(2017)\citenamefont
  {Kandala}, \citenamefont {Mezzacapo}, \citenamefont {Temme}, \citenamefont
  {Takita}, \citenamefont {Brink}, \citenamefont {Chow},\ and\ \citenamefont
  {Gambetta}}]{Kandala2017}%
  \BibitemOpen
  \bibfield  {author} {\bibinfo {author} {\bibfnamefont {A.}~\bibnamefont
  {Kandala}}, \bibinfo {author} {\bibfnamefont {A.}~\bibnamefont {Mezzacapo}},
  \bibinfo {author} {\bibfnamefont {K.}~\bibnamefont {Temme}}, \bibinfo
  {author} {\bibfnamefont {M.}~\bibnamefont {Takita}}, \bibinfo {author}
  {\bibfnamefont {M.}~\bibnamefont {Brink}}, \bibinfo {author} {\bibfnamefont
  {J.~M.}\ \bibnamefont {Chow}},\ and\ \bibinfo {author} {\bibfnamefont
  {J.~M.}\ \bibnamefont {Gambetta}},\ }\bibfield  {title} {\bibinfo {title}
  {Hardware-efficient variational quantum eigensolver for small molecules and
  quantum magnets},\ }\href {https://doi.org/10.1038/nature23879} {\bibfield
  {journal} {\bibinfo  {journal} {Nature}\ }\textbf {\bibinfo {volume} {549}},\
  \bibinfo {pages} {242} (\bibinfo {year} {2017})}\BibitemShut {NoStop}%
\bibitem [{\citenamefont {Browaeys}\ and\ \citenamefont
  {Lahaye}(2020)}]{browaeys_many-body_2020}%
  \BibitemOpen
  \bibfield  {author} {\bibinfo {author} {\bibfnamefont {A.}~\bibnamefont
  {Browaeys}}\ and\ \bibinfo {author} {\bibfnamefont {T.}~\bibnamefont
  {Lahaye}},\ }\bibfield  {title} {\bibinfo {title} {Many-body physics with
  individually controlled {Rydberg} atoms},\ }\href
  {https://doi.org/10.1038/s41567-019-0733-z} {\bibfield  {journal} {\bibinfo
  {journal} {Nature Physics}\ }\textbf {\bibinfo {volume} {16}},\ \bibinfo
  {pages} {132} (\bibinfo {year} {2020})}\BibitemShut {NoStop}%
\bibitem [{\citenamefont {Scholl}\ \emph {et~al.}(2021)\citenamefont {Scholl},
  \citenamefont {Schuler}, \citenamefont {Williams}, \citenamefont
  {Eberharter}, \citenamefont {Barredo}, \citenamefont {Schymik}, \citenamefont
  {Lienhard}, \citenamefont {Henry}, \citenamefont {Lang}, \citenamefont
  {Lahaye}, \citenamefont {Läuchli},\ and\ \citenamefont
  {Browaeys}}]{scholl_programmable_2021}%
  \BibitemOpen
  \bibfield  {author} {\bibinfo {author} {\bibfnamefont {P.}~\bibnamefont
  {Scholl}}, \bibinfo {author} {\bibfnamefont {M.}~\bibnamefont {Schuler}},
  \bibinfo {author} {\bibfnamefont {H.~J.}\ \bibnamefont {Williams}}, \bibinfo
  {author} {\bibfnamefont {A.~A.}\ \bibnamefont {Eberharter}}, \bibinfo
  {author} {\bibfnamefont {D.}~\bibnamefont {Barredo}}, \bibinfo {author}
  {\bibfnamefont {K.-N.}\ \bibnamefont {Schymik}}, \bibinfo {author}
  {\bibfnamefont {V.}~\bibnamefont {Lienhard}}, \bibinfo {author}
  {\bibfnamefont {L.-P.}\ \bibnamefont {Henry}}, \bibinfo {author}
  {\bibfnamefont {T.~C.}\ \bibnamefont {Lang}}, \bibinfo {author}
  {\bibfnamefont {T.}~\bibnamefont {Lahaye}}, \bibinfo {author} {\bibfnamefont
  {A.~M.}\ \bibnamefont {Läuchli}},\ and\ \bibinfo {author} {\bibfnamefont
  {A.}~\bibnamefont {Browaeys}},\ }\bibfield  {title} {\bibinfo {title}
  {Programmable quantum simulation of {2D} antiferromagnets with hundreds of
  {Rydberg} atoms},\ }\href {https://doi.org/10.1038/s41586-021-03585-1}
  {\bibfield  {journal} {\bibinfo  {journal} {Nature}\ }\textbf {\bibinfo
  {volume} {595}},\ \bibinfo {pages} {233} (\bibinfo {year} {2021})},\ \bibinfo
  {note} {arXiv: 2012.12268}\BibitemShut {NoStop}%
\bibitem [{\citenamefont {Ebadi}\ \emph {et~al.}(2022)\citenamefont {Ebadi}
  \emph {et~al.}}]{ebadi_quantum_2022}%
  \BibitemOpen
  \bibfield  {author} {\bibinfo {author} {\bibfnamefont {S.}~\bibnamefont
  {Ebadi}} \emph {et~al.},\ }\bibfield  {title} {\bibinfo {title} {Quantum
  optimization of maximum independent set using rydberg atom arrays},\ }\href
  {https://doi.org/10.1126/science.abo6587} {\bibfield  {journal} {\bibinfo
  {journal} {Science}\ }\textbf {\bibinfo {volume} {376}},\ \bibinfo {pages}
  {1209} (\bibinfo {year} {2022})}\BibitemShut {NoStop}%
\bibitem [{\citenamefont {Chen}\ \emph {et~al.}(2023)\citenamefont {Chen} \emph
  {et~al.}}]{chen_continuous_2023}%
  \BibitemOpen
  \bibfield  {author} {\bibinfo {author} {\bibfnamefont {C.}~\bibnamefont
  {Chen}} \emph {et~al.},\ }\bibfield  {title} {\bibinfo {title} {Continuous
  {Symmetry} {Breaking} in a {Two}-dimensional {Rydberg} {Array}},\ }\href
  {https://doi.org/10.1038/s41586-023-05859-2} {\bibfield  {journal} {\bibinfo
  {journal} {Nature}\ ,\ \bibinfo {pages} {1}} (\bibinfo {year}
  {2023})}\BibitemShut {NoStop}%
\bibitem [{\citenamefont {Henriet}\ \emph {et~al.}(2020)\citenamefont
  {Henriet}, \citenamefont {Beguin}, \citenamefont {Signoles}, \citenamefont
  {Lahaye}, \citenamefont {Browaeys}, \citenamefont {Reymond},\ and\
  \citenamefont {Jurczak}}]{henriet_quantum_2020}%
  \BibitemOpen
  \bibfield  {author} {\bibinfo {author} {\bibfnamefont {L.}~\bibnamefont
  {Henriet}}, \bibinfo {author} {\bibfnamefont {L.}~\bibnamefont {Beguin}},
  \bibinfo {author} {\bibfnamefont {A.}~\bibnamefont {Signoles}}, \bibinfo
  {author} {\bibfnamefont {T.}~\bibnamefont {Lahaye}}, \bibinfo {author}
  {\bibfnamefont {A.}~\bibnamefont {Browaeys}}, \bibinfo {author}
  {\bibfnamefont {G.-O.}\ \bibnamefont {Reymond}},\ and\ \bibinfo {author}
  {\bibfnamefont {C.}~\bibnamefont {Jurczak}},\ }\bibfield  {title} {\bibinfo
  {title} {Quantum computing with neutral atoms},\ }\href
  {https://doi.org/10.22331/q-2020-09-21-327} {\bibfield  {journal} {\bibinfo
  {journal} {Quantum}\ }\textbf {\bibinfo {volume} {4}},\ \bibinfo {pages}
  {327} (\bibinfo {year} {2020})},\ \bibinfo {note} {arXiv:
  2006.12326}\BibitemShut {NoStop}%
\bibitem [{\citenamefont {Wecker}\ \emph {et~al.}(2015)\citenamefont {Wecker},
  \citenamefont {Hastings},\ and\ \citenamefont
  {Troyer}}]{wecker_progress_2015}%
  \BibitemOpen
  \bibfield  {author} {\bibinfo {author} {\bibfnamefont {D.}~\bibnamefont
  {Wecker}}, \bibinfo {author} {\bibfnamefont {M.~B.}\ \bibnamefont
  {Hastings}},\ and\ \bibinfo {author} {\bibfnamefont {M.}~\bibnamefont
  {Troyer}},\ }\bibfield  {title} {\bibinfo {title} {Progress towards practical
  quantum variational algorithms},\ }\href
  {https://doi.org/10.1103/PhysRevA.92.042303} {\bibfield  {journal} {\bibinfo
  {journal} {Physical Review A}\ }\textbf {\bibinfo {volume} {92}},\ \bibinfo
  {pages} {042303} (\bibinfo {year} {2015})}\BibitemShut {NoStop}%
\bibitem [{\citenamefont {Cerezo}\ \emph {et~al.}(2021)\citenamefont {Cerezo},
  \citenamefont {Arrasmith}, \citenamefont {Babbush}, \citenamefont {Benjamin},
  \citenamefont {Endo}, \citenamefont {Fujii}, \citenamefont {McClean},
  \citenamefont {Mitarai}, \citenamefont {Yuan}, \citenamefont {Cincio},\ and\
  \citenamefont {Coles}}]{cerezo_variational_2021}%
  \BibitemOpen
  \bibfield  {author} {\bibinfo {author} {\bibfnamefont {M.}~\bibnamefont
  {Cerezo}}, \bibinfo {author} {\bibfnamefont {A.}~\bibnamefont {Arrasmith}},
  \bibinfo {author} {\bibfnamefont {R.}~\bibnamefont {Babbush}}, \bibinfo
  {author} {\bibfnamefont {S.~C.}\ \bibnamefont {Benjamin}}, \bibinfo {author}
  {\bibfnamefont {S.}~\bibnamefont {Endo}}, \bibinfo {author} {\bibfnamefont
  {K.}~\bibnamefont {Fujii}}, \bibinfo {author} {\bibfnamefont {J.~R.}\
  \bibnamefont {McClean}}, \bibinfo {author} {\bibfnamefont {K.}~\bibnamefont
  {Mitarai}}, \bibinfo {author} {\bibfnamefont {X.}~\bibnamefont {Yuan}},
  \bibinfo {author} {\bibfnamefont {L.}~\bibnamefont {Cincio}},\ and\ \bibinfo
  {author} {\bibfnamefont {P.~J.}\ \bibnamefont {Coles}},\ }\bibfield  {title}
  {\bibinfo {title} {Variational {Quantum} {Algorithms}},\ }\href
  {https://doi.org/10.1038/s42254-021-00348-9} {\bibfield  {journal} {\bibinfo
  {journal} {Nature Reviews Physics}\ }\textbf {\bibinfo {volume} {3}},\
  \bibinfo {pages} {625} (\bibinfo {year} {2021})},\ \bibinfo {note} {arXiv:
  2012.09265}\BibitemShut {NoStop}%
\bibitem [{\citenamefont {Barkoutsos}\ \emph {et~al.}(2020)\citenamefont
  {Barkoutsos}, \citenamefont {Nannicini}, \citenamefont {Robert},
  \citenamefont {Tavernelli},\ and\ \citenamefont
  {Woerner}}]{barkoutsos_improving_2020}%
  \BibitemOpen
  \bibfield  {author} {\bibinfo {author} {\bibfnamefont {P.~K.}\ \bibnamefont
  {Barkoutsos}}, \bibinfo {author} {\bibfnamefont {G.}~\bibnamefont
  {Nannicini}}, \bibinfo {author} {\bibfnamefont {A.}~\bibnamefont {Robert}},
  \bibinfo {author} {\bibfnamefont {I.}~\bibnamefont {Tavernelli}},\ and\
  \bibinfo {author} {\bibfnamefont {S.}~\bibnamefont {Woerner}},\ }\bibfield
  {title} {\bibinfo {title} {Improving {Variational} {Quantum} {Optimization}
  using {CVaR}},\ }\href {https://doi.org/10.22331/q-2020-04-20-256} {\bibfield
   {journal} {\bibinfo  {journal} {Quantum}\ }\textbf {\bibinfo {volume} {4}},\
  \bibinfo {pages} {256} (\bibinfo {year} {2020})},\ \bibinfo {note} {arXiv:
  1907.04769}\BibitemShut {NoStop}%
\bibitem [{\citenamefont {McClean}\ \emph {et~al.}(2018)\citenamefont
  {McClean}, \citenamefont {Boixo}, \citenamefont {Smelyanskiy}, \citenamefont
  {Babbush},\ and\ \citenamefont {Neven}}]{mcclean_barren_2018}%
  \BibitemOpen
  \bibfield  {author} {\bibinfo {author} {\bibfnamefont {J.~R.}\ \bibnamefont
  {McClean}}, \bibinfo {author} {\bibfnamefont {S.}~\bibnamefont {Boixo}},
  \bibinfo {author} {\bibfnamefont {V.~N.}\ \bibnamefont {Smelyanskiy}},
  \bibinfo {author} {\bibfnamefont {R.}~\bibnamefont {Babbush}},\ and\ \bibinfo
  {author} {\bibfnamefont {H.}~\bibnamefont {Neven}},\ }\bibfield  {title}
  {\bibinfo {title} {Barren plateaus in quantum neural network training
  landscapes},\ }\href {https://doi.org/10.1038/s41467-018-07090-4} {\bibfield
  {journal} {\bibinfo  {journal} {Nature Communications}\ }\textbf {\bibinfo
  {volume} {9}},\ \bibinfo {pages} {4812} (\bibinfo {year} {2018})}\BibitemShut
  {NoStop}%
\bibitem [{\citenamefont {Meitei}\ \emph {et~al.}(2021)\citenamefont {Meitei},
  \citenamefont {Gard}, \citenamefont {Barron}, \citenamefont {Pappas},
  \citenamefont {Economou}, \citenamefont {Barnes},\ and\ \citenamefont
  {Mayhall}}]{meitei_gate-free_2021}%
  \BibitemOpen
  \bibfield  {author} {\bibinfo {author} {\bibfnamefont {O.~R.}\ \bibnamefont
  {Meitei}}, \bibinfo {author} {\bibfnamefont {B.~T.}\ \bibnamefont {Gard}},
  \bibinfo {author} {\bibfnamefont {G.~S.}\ \bibnamefont {Barron}}, \bibinfo
  {author} {\bibfnamefont {D.~P.}\ \bibnamefont {Pappas}}, \bibinfo {author}
  {\bibfnamefont {S.~E.}\ \bibnamefont {Economou}}, \bibinfo {author}
  {\bibfnamefont {E.}~\bibnamefont {Barnes}},\ and\ \bibinfo {author}
  {\bibfnamefont {N.~J.}\ \bibnamefont {Mayhall}},\ }\bibfield  {title}
  {\bibinfo {title} {Gate-free state preparation for fast variational quantum
  eigensolver simulations},\ }\href
  {https://doi.org/10.1038/s41534-021-00493-0} {\bibfield  {journal} {\bibinfo
  {journal} {npj Quantum Information}\ }\textbf {\bibinfo {volume} {7}},\
  \bibinfo {pages} {155} (\bibinfo {year} {2021})}\BibitemShut {NoStop}%
\bibitem [{\citenamefont {Wakaura}\ \emph {et~al.}(2021)\citenamefont
  {Wakaura}, \citenamefont {Tomono},\ and\ \citenamefont
  {Yasuda}}]{wakaura_evaluation_2021}%
  \BibitemOpen
  \bibfield  {author} {\bibinfo {author} {\bibfnamefont {H.}~\bibnamefont
  {Wakaura}}, \bibinfo {author} {\bibfnamefont {T.}~\bibnamefont {Tomono}},\
  and\ \bibinfo {author} {\bibfnamefont {S.}~\bibnamefont {Yasuda}},\
  }\bibfield  {title} {\bibinfo {title} {Evaluation on {Genetic} {Algorithms}
  as an optimizer of {Variational} {Quantum} {Eigensolver}({VQE}) method},\
  }\href {http://arxiv.org/abs/2110.07441} {\bibfield  {journal} {\bibinfo
  {journal} {arXiv:2110.07441 [quant-ph]}\ } (\bibinfo {year} {2021})},\
  \bibinfo {note} {arXiv: 2110.07441}\BibitemShut {NoStop}%
\bibitem [{\citenamefont {Banchi}\ and\ \citenamefont
  {Crooks}(2021)}]{banchi_measuring_2021}%
  \BibitemOpen
  \bibfield  {author} {\bibinfo {author} {\bibfnamefont {L.}~\bibnamefont
  {Banchi}}\ and\ \bibinfo {author} {\bibfnamefont {G.~E.}\ \bibnamefont
  {Crooks}},\ }\bibfield  {title} {\bibinfo {title} {Measuring {Analytic}
  {Gradients} of {General} {Quantum} {Evolution} with the {Stochastic}
  {Parameter} {Shift} {Rule}},\ }\href
  {https://doi.org/10.22331/q-2021-01-25-386} {\bibfield  {journal} {\bibinfo
  {journal} {Quantum}\ }\textbf {\bibinfo {volume} {5}},\ \bibinfo {pages}
  {386} (\bibinfo {year} {2021})},\ \bibinfo {note} {arXiv:
  2005.10299}\BibitemShut {NoStop}%
\bibitem [{\citenamefont {Gacon}\ \emph {et~al.}(2021)\citenamefont {Gacon},
  \citenamefont {Zoufal}, \citenamefont {Carleo},\ and\ \citenamefont
  {Woerner}}]{gacon2021simultaneous}%
  \BibitemOpen
  \bibfield  {author} {\bibinfo {author} {\bibfnamefont {J.}~\bibnamefont
  {Gacon}}, \bibinfo {author} {\bibfnamefont {C.}~\bibnamefont {Zoufal}},
  \bibinfo {author} {\bibfnamefont {G.}~\bibnamefont {Carleo}},\ and\ \bibinfo
  {author} {\bibfnamefont {S.}~\bibnamefont {Woerner}},\ }\bibfield  {title}
  {\bibinfo {title} {Simultaneous perturbation stochastic approximation of the
  quantum fisher information},\ }\href
  {https://doi.org/10.22331/q-2021-10-20-567} {\bibfield  {journal} {\bibinfo
  {journal} {Quantum}\ }\textbf {\bibinfo {volume} {5}},\ \bibinfo {pages}
  {567} (\bibinfo {year} {2021})}\BibitemShut {NoStop}%
\bibitem [{\citenamefont {Piskor}\ \emph {et~al.}(2022)\citenamefont {Piskor},
  \citenamefont {Reiner}, \citenamefont {Zanker}, \citenamefont {Vogt},
  \citenamefont {Marthaler}, \citenamefont {Wilhelm},\ and\ \citenamefont
  {Eich}}]{piskor_using_2022}%
  \BibitemOpen
  \bibfield  {author} {\bibinfo {author} {\bibfnamefont {T.}~\bibnamefont
  {Piskor}}, \bibinfo {author} {\bibfnamefont {J.-M.}\ \bibnamefont {Reiner}},
  \bibinfo {author} {\bibfnamefont {S.}~\bibnamefont {Zanker}}, \bibinfo
  {author} {\bibfnamefont {N.}~\bibnamefont {Vogt}}, \bibinfo {author}
  {\bibfnamefont {M.}~\bibnamefont {Marthaler}}, \bibinfo {author}
  {\bibfnamefont {F.~K.}\ \bibnamefont {Wilhelm}},\ and\ \bibinfo {author}
  {\bibfnamefont {F.~G.}\ \bibnamefont {Eich}},\ }\bibfield  {title} {\bibinfo
  {title} {Using gradient-based algorithms to determine ground-state energies
  on a quantum computer},\ }\href {https://doi.org/10.1103/PhysRevA.105.062415}
  {\bibfield  {journal} {\bibinfo  {journal} {Physical Review A}\ }\textbf
  {\bibinfo {volume} {105}},\ \bibinfo {pages} {062415} (\bibinfo {year}
  {2022})}\BibitemShut {NoStop}%
\bibitem [{\citenamefont {Huang}\ \emph {et~al.}(2021)\citenamefont {Huang},
  \citenamefont {Kueng},\ and\ \citenamefont
  {Preskill}}]{huang_efficient_2021}%
  \BibitemOpen
  \bibfield  {author} {\bibinfo {author} {\bibfnamefont {H.-Y.}\ \bibnamefont
  {Huang}}, \bibinfo {author} {\bibfnamefont {R.}~\bibnamefont {Kueng}},\ and\
  \bibinfo {author} {\bibfnamefont {J.}~\bibnamefont {Preskill}},\ }\bibfield
  {title} {\bibinfo {title} {Efficient estimation of {Pauli} observables by
  derandomization},\ }\href {https://doi.org/10.1103/PhysRevLett.127.030503}
  {\bibfield  {journal} {\bibinfo  {journal} {Physical Review Letters}\
  }\textbf {\bibinfo {volume} {127}},\ \bibinfo {pages} {030503} (\bibinfo
  {year} {2021})},\ \bibinfo {note} {arXiv: 2103.07510}\BibitemShut {NoStop}%
\bibitem [{\citenamefont {Elben}\ \emph {et~al.}(2019)\citenamefont {Elben},
  \citenamefont {Vermersch}, \citenamefont {Roos},\ and\ \citenamefont
  {Zoller}}]{elben_statistical_2019}%
  \BibitemOpen
  \bibfield  {author} {\bibinfo {author} {\bibfnamefont {A.}~\bibnamefont
  {Elben}}, \bibinfo {author} {\bibfnamefont {B.}~\bibnamefont {Vermersch}},
  \bibinfo {author} {\bibfnamefont {C.~F.}\ \bibnamefont {Roos}},\ and\
  \bibinfo {author} {\bibfnamefont {P.}~\bibnamefont {Zoller}},\ }\bibfield
  {title} {\bibinfo {title} {Statistical correlations between locally
  randomized measurements: {A} toolbox for probing entanglement in many-body
  quantum states},\ }\href {https://doi.org/10.1103/PhysRevA.99.052323}
  {\bibfield  {journal} {\bibinfo  {journal} {Physical Review A}\ }\textbf
  {\bibinfo {volume} {99}},\ \bibinfo {pages} {052323} (\bibinfo {year}
  {2019})}\BibitemShut {NoStop}%
\bibitem [{\citenamefont {Kokail}\ \emph {et~al.}(2019)\citenamefont {Kokail},
  \citenamefont {Maier}, \citenamefont {van Bijnen}, \citenamefont {Brydges},
  \citenamefont {Joshi}, \citenamefont {Jurcevic}, \citenamefont {Muschik},
  \citenamefont {Silvi}, \citenamefont {Blatt}, \citenamefont {Roos},\ and\
  \citenamefont {Zoller}}]{kokail_self-verifying_2019}%
  \BibitemOpen
  \bibfield  {author} {\bibinfo {author} {\bibfnamefont {C.}~\bibnamefont
  {Kokail}}, \bibinfo {author} {\bibfnamefont {C.}~\bibnamefont {Maier}},
  \bibinfo {author} {\bibfnamefont {R.}~\bibnamefont {van Bijnen}}, \bibinfo
  {author} {\bibfnamefont {T.}~\bibnamefont {Brydges}}, \bibinfo {author}
  {\bibfnamefont {M.~K.}\ \bibnamefont {Joshi}}, \bibinfo {author}
  {\bibfnamefont {P.}~\bibnamefont {Jurcevic}}, \bibinfo {author}
  {\bibfnamefont {C.~A.}\ \bibnamefont {Muschik}}, \bibinfo {author}
  {\bibfnamefont {P.}~\bibnamefont {Silvi}}, \bibinfo {author} {\bibfnamefont
  {R.}~\bibnamefont {Blatt}}, \bibinfo {author} {\bibfnamefont {C.~F.}\
  \bibnamefont {Roos}},\ and\ \bibinfo {author} {\bibfnamefont
  {P.}~\bibnamefont {Zoller}},\ }\bibfield  {title} {\bibinfo {title}
  {Self-{Verifying} {Variational} {Quantum} {Simulation} of the {Lattice}
  {Schwinger} {Model}},\ }\href {https://doi.org/10.1038/s41586-019-1177-4}
  {\bibfield  {journal} {\bibinfo  {journal} {Nature}\ }\textbf {\bibinfo
  {volume} {569}},\ \bibinfo {pages} {355} (\bibinfo {year} {2019})},\ \bibinfo
  {note} {arXiv: 1810.03421}\BibitemShut {NoStop}%
\bibitem [{\citenamefont {Nam}\ \emph {et~al.}(2020)\citenamefont {Nam} \emph
  {et~al.}}]{nam_ground-state_2020}%
  \BibitemOpen
  \bibfield  {author} {\bibinfo {author} {\bibfnamefont {Y.}~\bibnamefont
  {Nam}} \emph {et~al.},\ }\bibfield  {title} {\bibinfo {title} {Ground-state
  energy estimation of the water molecule on a trapped-ion quantum computer},\
  }\href {https://doi.org/10.1038/s41534-020-0259-3} {\bibfield  {journal}
  {\bibinfo  {journal} {npj Quantum Information}\ }\textbf {\bibinfo {volume}
  {6}},\ \bibinfo {pages} {33} (\bibinfo {year} {2020})}\BibitemShut {NoStop}%
\bibitem [{\citenamefont {Dalyac}\ \emph {et~al.}(2021)\citenamefont {Dalyac},
  \citenamefont {Henriet}, \citenamefont {Jeandel}, \citenamefont {Lechner},
  \citenamefont {Perdrix}, \citenamefont {Porcheron},\ and\ \citenamefont
  {Veshchezerova}}]{dalyac_qualifying_2021}%
  \BibitemOpen
  \bibfield  {author} {\bibinfo {author} {\bibfnamefont {C.}~\bibnamefont
  {Dalyac}}, \bibinfo {author} {\bibfnamefont {L.}~\bibnamefont {Henriet}},
  \bibinfo {author} {\bibfnamefont {E.}~\bibnamefont {Jeandel}}, \bibinfo
  {author} {\bibfnamefont {W.}~\bibnamefont {Lechner}}, \bibinfo {author}
  {\bibfnamefont {S.}~\bibnamefont {Perdrix}}, \bibinfo {author} {\bibfnamefont
  {M.}~\bibnamefont {Porcheron}},\ and\ \bibinfo {author} {\bibfnamefont
  {M.}~\bibnamefont {Veshchezerova}},\ }\bibfield  {title} {\bibinfo {title}
  {Qualifying quantum approaches for hard industrial optimization problems. {A}
  case study in the field of smart-charging of electric vehicles},\ }\href
  {https://doi.org/10.1140/epjqt/s40507-021-00100-3} {\bibfield  {journal}
  {\bibinfo  {journal} {EPJ Quantum Technology}\ }\textbf {\bibinfo {volume}
  {8}},\ \bibinfo {pages} {1} (\bibinfo {year} {2021})}\BibitemShut {NoStop}%
\bibitem [{\citenamefont {Parra-Rodriguez}\ \emph {et~al.}(2020)\citenamefont
  {Parra-Rodriguez}, \citenamefont {Lougovski}, \citenamefont {Lamata},
  \citenamefont {Solano},\ and\ \citenamefont
  {Sanz}}]{parra-rodriguez_digital-analog_2020}%
  \BibitemOpen
  \bibfield  {author} {\bibinfo {author} {\bibfnamefont {A.}~\bibnamefont
  {Parra-Rodriguez}}, \bibinfo {author} {\bibfnamefont {P.}~\bibnamefont
  {Lougovski}}, \bibinfo {author} {\bibfnamefont {L.}~\bibnamefont {Lamata}},
  \bibinfo {author} {\bibfnamefont {E.}~\bibnamefont {Solano}},\ and\ \bibinfo
  {author} {\bibfnamefont {M.}~\bibnamefont {Sanz}},\ }\bibfield  {title}
  {\bibinfo {title} {Digital-analog quantum computation},\ }\href
  {https://doi.org/10.1103/PhysRevA.101.022305} {\bibfield  {journal} {\bibinfo
   {journal} {Physical Review A}\ }\textbf {\bibinfo {volume} {101}},\ \bibinfo
  {pages} {022305} (\bibinfo {year} {2020})}\BibitemShut {NoStop}%
\bibitem [{\citenamefont {Fedorov}\ \emph {et~al.}(2022)\citenamefont
  {Fedorov}, \citenamefont {Peng}, \citenamefont {Govind},\ and\ \citenamefont
  {Alexeev}}]{fedorov_vqe_2022}%
  \BibitemOpen
  \bibfield  {author} {\bibinfo {author} {\bibfnamefont {D.~A.}\ \bibnamefont
  {Fedorov}}, \bibinfo {author} {\bibfnamefont {B.}~\bibnamefont {Peng}},
  \bibinfo {author} {\bibfnamefont {N.}~\bibnamefont {Govind}},\ and\ \bibinfo
  {author} {\bibfnamefont {Y.}~\bibnamefont {Alexeev}},\ }\bibfield  {title}
  {\bibinfo {title} {{VQE} method: a short survey and recent developments},\
  }\href {https://doi.org/10.1186/s41313-021-00032-6} {\bibfield  {journal}
  {\bibinfo  {journal} {Materials Theory}\ }\textbf {\bibinfo {volume} {6}},\
  \bibinfo {pages} {2} (\bibinfo {year} {2022})}\BibitemShut {NoStop}%
\bibitem [{\citenamefont {Jordan}\ and\ \citenamefont
  {Wigner}(1928)}]{jordan_uber_1928}%
  \BibitemOpen
  \bibfield  {author} {\bibinfo {author} {\bibfnamefont {P.}~\bibnamefont
  {Jordan}}\ and\ \bibinfo {author} {\bibfnamefont {E.}~\bibnamefont
  {Wigner}},\ }\bibfield  {title} {\bibinfo {title} {Über das {Paulische}
  Äquivalenzverbot},\ }\href {https://doi.org/10.1007/BF01331938} {\bibfield
  {journal} {\bibinfo  {journal} {Zeitschrift für Physik}\ }\textbf {\bibinfo
  {volume} {47}},\ \bibinfo {pages} {631} (\bibinfo {year} {1928})}\BibitemShut
  {NoStop}%
\bibitem [{\citenamefont {Bravyi}\ and\ \citenamefont
  {Kitaev}(2002)}]{bravyi_fermionic_2002}%
  \BibitemOpen
  \bibfield  {author} {\bibinfo {author} {\bibfnamefont {S.}~\bibnamefont
  {Bravyi}}\ and\ \bibinfo {author} {\bibfnamefont {A.}~\bibnamefont
  {Kitaev}},\ }\bibfield  {title} {\bibinfo {title} {Fermionic quantum
  computation},\ }\href {https://doi.org/10.1006/aphy.2002.6254} {\bibfield
  {journal} {\bibinfo  {journal} {Annals of Physics}\ }\textbf {\bibinfo
  {volume} {298}},\ \bibinfo {pages} {210} (\bibinfo {year} {2002})},\ \bibinfo
  {note} {arXiv: quant-ph/0003137}\BibitemShut {NoStop}%
\bibitem [{\citenamefont {Bartlett}\ \emph {et~al.}(1989)\citenamefont
  {Bartlett}, \citenamefont {Kucharski},\ and\ \citenamefont
  {Noga}}]{bartlett_alternative_1989}%
  \BibitemOpen
  \bibfield  {author} {\bibinfo {author} {\bibfnamefont {R.~J.}\ \bibnamefont
  {Bartlett}}, \bibinfo {author} {\bibfnamefont {S.~A.}\ \bibnamefont
  {Kucharski}},\ and\ \bibinfo {author} {\bibfnamefont {J.}~\bibnamefont
  {Noga}},\ }\bibfield  {title} {\bibinfo {title} {Alternative coupled-cluster
  ansätze {II}. {The} unitary coupled-cluster method},\ }\href
  {https://doi.org/10.1016/S0009-2614(89)87372-5} {\bibfield  {journal}
  {\bibinfo  {journal} {Chemical Physics Letters}\ }\textbf {\bibinfo {volume}
  {155}},\ \bibinfo {pages} {133} (\bibinfo {year} {1989})}\BibitemShut
  {NoStop}%
\bibitem [{\citenamefont {O'Malley}\ \emph {et~al.}(2016)\citenamefont
  {O'Malley} \emph {et~al.}}]{omalley_scalable_2016}%
  \BibitemOpen
  \bibfield  {author} {\bibinfo {author} {\bibfnamefont {P.~J.~J.}\
  \bibnamefont {O'Malley}} \emph {et~al.},\ }\bibfield  {title} {\bibinfo
  {title} {Scalable quantum simulation of molecular energies},\ }\href
  {https://doi.org/10.1103/PhysRevX.6.031007} {\bibfield  {journal} {\bibinfo
  {journal} {Phys. Rev. X}\ }\textbf {\bibinfo {volume} {6}},\ \bibinfo {pages}
  {031007} (\bibinfo {year} {2016})}\BibitemShut {NoStop}%
\bibitem [{\citenamefont {Helgaker}\ \emph {et~al.}(2000)\citenamefont
  {Helgaker}, \citenamefont {Jørgensen},\ and\ \citenamefont
  {Olsen}}]{helgaker_molecular_2014}%
  \BibitemOpen
  \bibfield  {author} {\bibinfo {author} {\bibfnamefont {T.}~\bibnamefont
  {Helgaker}}, \bibinfo {author} {\bibfnamefont {P.}~\bibnamefont
  {Jørgensen}},\ and\ \bibinfo {author} {\bibfnamefont {J.}~\bibnamefont
  {Olsen}},\ }\bibinfo {title} {Coupled-cluster theory},\ in\ \href
  {https://doi.org/https://doi.org/10.1002/9781119019572.ch13} {\emph {\bibinfo
  {booktitle} {Molecular Electronic‐Structure Theory}}}\ (\bibinfo
  {publisher} {John Wiley \& Sons, Ltd},\ \bibinfo {year} {2000})\
  Chap.~\bibinfo {chapter} {13}, pp.\ \bibinfo {pages} {648--723}\BibitemShut
  {NoStop}%
\bibitem [{\citenamefont {Storn}\ and\ \citenamefont
  {Price}(1997)}]{storn_differential_1997}%
  \BibitemOpen
  \bibfield  {author} {\bibinfo {author} {\bibfnamefont {R.}~\bibnamefont
  {Storn}}\ and\ \bibinfo {author} {\bibfnamefont {K.}~\bibnamefont {Price}},\
  }\bibfield  {title} {\bibinfo {title} {Differential {Evolution} – {A}
  {Simple} and {Efficient} {Heuristic} for global {Optimization} over
  {Continuous} {Spaces}},\ }\href {https://doi.org/10.1023/A:1008202821328}
  {\bibfield  {journal} {\bibinfo  {journal} {Journal of Global Optimization}\
  }\textbf {\bibinfo {volume} {11}},\ \bibinfo {pages} {341} (\bibinfo {year}
  {1997})}\BibitemShut {NoStop}%
\bibitem [{\citenamefont {Farhi}\ \emph {et~al.}(2014)\citenamefont {Farhi},
  \citenamefont {Goldstone},\ and\ \citenamefont
  {Gutmann}}]{farhi_quantum_2014}%
  \BibitemOpen
  \bibfield  {author} {\bibinfo {author} {\bibfnamefont {E.}~\bibnamefont
  {Farhi}}, \bibinfo {author} {\bibfnamefont {J.}~\bibnamefont {Goldstone}},\
  and\ \bibinfo {author} {\bibfnamefont {S.}~\bibnamefont {Gutmann}},\
  }\bibfield  {title} {\bibinfo {title} {A {Quantum} {Approximate}
  {Optimization} {Algorithm}},\ }\href {http://arxiv.org/abs/1411.4028}
  {\bibfield  {journal} {\bibinfo  {journal} {arXiv:1411.4028}\ } (\bibinfo
  {year} {2014})}\BibitemShut {NoStop}%
\bibitem [{\citenamefont {Leclerc}\ \emph {et~al.}(2022)\citenamefont
  {Leclerc}, \citenamefont {Ortiz-Guitierrez}, \citenamefont {Grijalva},
  \citenamefont {Albrecht}, \citenamefont {Cline}, \citenamefont {Elfving},
  \citenamefont {Signoles}, \citenamefont {Henriet}, \citenamefont {Del~Bimbo},
  \citenamefont {Sheikh}, \citenamefont {Shah}, \citenamefont {Andrea},
  \citenamefont {Ishtiaq}, \citenamefont {Duarte}, \citenamefont {Mugel},
  \citenamefont {Caceres}, \citenamefont {Kurek}, \citenamefont {Orus},
  \citenamefont {Seddik}, \citenamefont {Hammammi}, \citenamefont {Isselnane},\
  and\ \citenamefont {M'tamon}}]{leclerc2022financial}%
  \BibitemOpen
  \bibfield  {author} {\bibinfo {author} {\bibfnamefont {L.}~\bibnamefont
  {Leclerc}}, \bibinfo {author} {\bibfnamefont {L.}~\bibnamefont
  {Ortiz-Guitierrez}}, \bibinfo {author} {\bibfnamefont {S.}~\bibnamefont
  {Grijalva}}, \bibinfo {author} {\bibfnamefont {B.}~\bibnamefont {Albrecht}},
  \bibinfo {author} {\bibfnamefont {J.~R.~K.}\ \bibnamefont {Cline}}, \bibinfo
  {author} {\bibfnamefont {V.~E.}\ \bibnamefont {Elfving}}, \bibinfo {author}
  {\bibfnamefont {A.}~\bibnamefont {Signoles}}, \bibinfo {author}
  {\bibfnamefont {L.}~\bibnamefont {Henriet}}, \bibinfo {author} {\bibfnamefont
  {G.}~\bibnamefont {Del~Bimbo}}, \bibinfo {author} {\bibfnamefont {U.~A.}\
  \bibnamefont {Sheikh}}, \bibinfo {author} {\bibfnamefont {M.}~\bibnamefont
  {Shah}}, \bibinfo {author} {\bibfnamefont {L.}~\bibnamefont {Andrea}},
  \bibinfo {author} {\bibfnamefont {F.}~\bibnamefont {Ishtiaq}}, \bibinfo
  {author} {\bibfnamefont {A.}~\bibnamefont {Duarte}}, \bibinfo {author}
  {\bibfnamefont {S.}~\bibnamefont {Mugel}}, \bibinfo {author} {\bibfnamefont
  {I.}~\bibnamefont {Caceres}}, \bibinfo {author} {\bibfnamefont
  {M.}~\bibnamefont {Kurek}}, \bibinfo {author} {\bibfnamefont
  {R.}~\bibnamefont {Orus}}, \bibinfo {author} {\bibfnamefont {A.}~\bibnamefont
  {Seddik}}, \bibinfo {author} {\bibfnamefont {O.}~\bibnamefont {Hammammi}},
  \bibinfo {author} {\bibfnamefont {H.}~\bibnamefont {Isselnane}},\ and\
  \bibinfo {author} {\bibfnamefont {D.}~\bibnamefont {M'tamon}},\ }\bibfield
  {title} {\bibinfo {title} {Financial risk management on a neutral atom
  quantum processor},\ }\href {https://arxiv.org/abs/2212.03223} {\bibfield
  {journal} {\bibinfo  {journal} {arXiv:2212.03223}\ } (\bibinfo {year}
  {2022})}\BibitemShut {NoStop}%
\bibitem [{\citenamefont {Coelho}\ \emph {et~al.}(2022)\citenamefont {Coelho},
  \citenamefont {D'Arcangelo},\ and\ \citenamefont
  {Henry}}]{coelho2022efficient}%
  \BibitemOpen
  \bibfield  {author} {\bibinfo {author} {\bibfnamefont {W.~d.~S.}\
  \bibnamefont {Coelho}}, \bibinfo {author} {\bibfnamefont {M.}~\bibnamefont
  {D'Arcangelo}},\ and\ \bibinfo {author} {\bibfnamefont {L.-P.}\ \bibnamefont
  {Henry}},\ }\bibfield  {title} {\bibinfo {title} {Efficient protocol for
  solving combinatorial graph problems on neutral-atom quantum processors},\
  }\href {https://arxiv.org/abs/2207.13030} {\bibfield  {journal} {\bibinfo
  {journal} {arXiv:2207.13030}\ } (\bibinfo {year} {2022})}\BibitemShut
  {NoStop}%
\bibitem [{\citenamefont {Notarnicola}\ \emph {et~al.}(2021)\citenamefont
  {Notarnicola}, \citenamefont {Elben}, \citenamefont {Lahaye}, \citenamefont
  {Browaeys}, \citenamefont {Montangero},\ and\ \citenamefont
  {Vermersch}}]{notarnicola_randomized_2021}%
  \BibitemOpen
  \bibfield  {author} {\bibinfo {author} {\bibfnamefont {S.}~\bibnamefont
  {Notarnicola}}, \bibinfo {author} {\bibfnamefont {A.}~\bibnamefont {Elben}},
  \bibinfo {author} {\bibfnamefont {T.}~\bibnamefont {Lahaye}}, \bibinfo
  {author} {\bibfnamefont {A.}~\bibnamefont {Browaeys}}, \bibinfo {author}
  {\bibfnamefont {S.}~\bibnamefont {Montangero}},\ and\ \bibinfo {author}
  {\bibfnamefont {B.}~\bibnamefont {Vermersch}},\ }\bibfield  {title} {\bibinfo
  {title} {A randomized measurement toolbox for {Rydberg} quantum
  technologies},\ }\href {http://arxiv.org/abs/2112.11046} {\bibfield
  {journal} {\bibinfo  {journal} {arXiv:2112.11046}\ } (\bibinfo {year}
  {2021})}\BibitemShut {NoStop}%
\bibitem [{\citenamefont {{Qiskit contributors}}(2023)}]{Qiskit}%
  \BibitemOpen
  \bibfield  {author} {\bibinfo {author} {\bibnamefont {{Qiskit
  contributors}}},\ }\href {https://doi.org/10.5281/zenodo.2573505} {\bibinfo
  {title} {Qiskit: An open-source framework for quantum computing}} (\bibinfo
  {year} {2023})\BibitemShut {NoStop}%
\bibitem [{\citenamefont {Muller}(2022)}]{muller_pyquante2_2022}%
  \BibitemOpen
  \bibfield  {author} {\bibinfo {author} {\bibfnamefont {R.}~\bibnamefont
  {Muller}},\ }\href {https://github.com/rpmuller/pyquante2} {\bibinfo {title}
  {{PyQuante2}}} (\bibinfo {year} {2022})\BibitemShut {NoStop}%
\bibitem [{\citenamefont {Silv\'erio}\ \emph {et~al.}(2022)\citenamefont
  {Silv\'erio}, \citenamefont {Grijalva}, \citenamefont {Dalyac}, \citenamefont
  {Leclerc}, \citenamefont {Karalekas}, \citenamefont {Shammah}, \citenamefont
  {Beji}, \citenamefont {Henry},\ and\ \citenamefont
  {Henriet}}]{silverio2022pulser}%
  \BibitemOpen
  \bibfield  {author} {\bibinfo {author} {\bibfnamefont {H.}~\bibnamefont
  {Silv\'erio}}, \bibinfo {author} {\bibfnamefont {S.}~\bibnamefont
  {Grijalva}}, \bibinfo {author} {\bibfnamefont {C.}~\bibnamefont {Dalyac}},
  \bibinfo {author} {\bibfnamefont {L.}~\bibnamefont {Leclerc}}, \bibinfo
  {author} {\bibfnamefont {P.~J.}\ \bibnamefont {Karalekas}}, \bibinfo {author}
  {\bibfnamefont {N.}~\bibnamefont {Shammah}}, \bibinfo {author} {\bibfnamefont
  {M.}~\bibnamefont {Beji}}, \bibinfo {author} {\bibfnamefont {L.-P.}\
  \bibnamefont {Henry}},\ and\ \bibinfo {author} {\bibfnamefont
  {L.}~\bibnamefont {Henriet}},\ }\bibfield  {title} {\bibinfo {title} {Pulser:
  An open-source package for the design of pulse sequences in programmable
  neutral-atom arrays},\ }\href {https://doi.org/10.22331/q-2022-01-24-629}
  {\bibfield  {journal} {\bibinfo  {journal} {Quantum}\ }\textbf {\bibinfo
  {volume} {6}},\ \bibinfo {pages} {629} (\bibinfo {year} {2022})}\BibitemShut
  {NoStop}%
\bibitem [{\citenamefont {Powell}(1964)}]{powell_efficient_1964}%
  \BibitemOpen
  \bibfield  {author} {\bibinfo {author} {\bibfnamefont {M.~J.~D.}\
  \bibnamefont {Powell}},\ }\bibfield  {title} {\bibinfo {title} {An efficient
  method for finding the minimum of a function of several variables without
  calculating derivatives},\ }\href {https://doi.org/10.1093/comjnl/7.2.155}
  {\bibfield  {journal} {\bibinfo  {journal} {The Computer Journal}\ }\textbf
  {\bibinfo {volume} {7}},\ \bibinfo {pages} {155} (\bibinfo {year}
  {1964})}\BibitemShut {NoStop}%
\bibitem [{\citenamefont {Nelder}\ and\ \citenamefont
  {Mead}(1965)}]{nelder_simplex_1965}%
  \BibitemOpen
  \bibfield  {author} {\bibinfo {author} {\bibfnamefont {J.~A.}\ \bibnamefont
  {Nelder}}\ and\ \bibinfo {author} {\bibfnamefont {R.}~\bibnamefont {Mead}},\
  }\bibfield  {title} {\bibinfo {title} {A {Simplex} {Method} for {Function}
  {Minimization}},\ }\href {https://doi.org/10.1093/comjnl/7.4.308} {\bibfield
  {journal} {\bibinfo  {journal} {The Computer Journal}\ }\textbf {\bibinfo
  {volume} {7}},\ \bibinfo {pages} {308} (\bibinfo {year} {1965})}\BibitemShut
  {NoStop}%
\bibitem [{\citenamefont {Schymik}\ \emph {et~al.}(2022)\citenamefont
  {Schymik}, \citenamefont {Ximenez}, \citenamefont {Bloch}, \citenamefont
  {Dreon}, \citenamefont {Signoles}, \citenamefont {Nogrette}, \citenamefont
  {Barredo}, \citenamefont {Browaeys},\ and\ \citenamefont
  {Lahaye}}]{schymik2022situ}%
  \BibitemOpen
  \bibfield  {author} {\bibinfo {author} {\bibfnamefont {K.-N.}\ \bibnamefont
  {Schymik}}, \bibinfo {author} {\bibfnamefont {B.}~\bibnamefont {Ximenez}},
  \bibinfo {author} {\bibfnamefont {E.}~\bibnamefont {Bloch}}, \bibinfo
  {author} {\bibfnamefont {D.}~\bibnamefont {Dreon}}, \bibinfo {author}
  {\bibfnamefont {A.}~\bibnamefont {Signoles}}, \bibinfo {author}
  {\bibfnamefont {F.}~\bibnamefont {Nogrette}}, \bibinfo {author}
  {\bibfnamefont {D.}~\bibnamefont {Barredo}}, \bibinfo {author} {\bibfnamefont
  {A.}~\bibnamefont {Browaeys}},\ and\ \bibinfo {author} {\bibfnamefont
  {T.}~\bibnamefont {Lahaye}},\ }\bibfield  {title} {\bibinfo {title} {In situ
  equalization of single-atom loading in large-scale optical tweezer arrays},\
  }\href {https://doi.org/10.1103} {\bibfield  {journal} {\bibinfo  {journal}
  {Physical Review A}\ }\textbf {\bibinfo {volume} {106}},\ \bibinfo {pages}
  {022611} (\bibinfo {year} {2022})}\BibitemShut {NoStop}%
\bibitem [{\citenamefont {Barredo}\ \emph {et~al.}(2018)\citenamefont
  {Barredo}, \citenamefont {Lienhard}, \citenamefont {de~L{\'{e}}s{\'{e}}leuc},
  \citenamefont {Lahaye},\ and\ \citenamefont {Browaeys}}]{barredo2018}%
  \BibitemOpen
  \bibfield  {author} {\bibinfo {author} {\bibfnamefont {D.}~\bibnamefont
  {Barredo}}, \bibinfo {author} {\bibfnamefont {V.}~\bibnamefont {Lienhard}},
  \bibinfo {author} {\bibfnamefont {S.}~\bibnamefont
  {de~L{\'{e}}s{\'{e}}leuc}}, \bibinfo {author} {\bibfnamefont
  {T.}~\bibnamefont {Lahaye}},\ and\ \bibinfo {author} {\bibfnamefont
  {A.}~\bibnamefont {Browaeys}},\ }\bibfield  {title} {\bibinfo {title}
  {Synthetic three-dimensional atomic structures assembled atom by atom},\
  }\href {https://doi.org/10.1038/s41586-018-0450-2} {\bibfield  {journal}
  {\bibinfo  {journal} {Nature}\ }\textbf {\bibinfo {volume} {561}},\ \bibinfo
  {pages} {79} (\bibinfo {year} {2018})}\BibitemShut {NoStop}%
\bibitem [{\citenamefont {Dalyac}\ and\ \citenamefont
  {Henriet}(2022)}]{dalyac2022embedding}%
  \BibitemOpen
  \bibfield  {author} {\bibinfo {author} {\bibfnamefont {C.}~\bibnamefont
  {Dalyac}}\ and\ \bibinfo {author} {\bibfnamefont {L.}~\bibnamefont
  {Henriet}},\ }\bibfield  {title} {\bibinfo {title} {Embedding the mis problem
  for non-local graphs with bounded degree using 3d arrays of atoms},\ }\href
  {https://arxiv.org/pdf/2209.05164} {\bibfield  {journal} {\bibinfo  {journal}
  {arXiv:2209.05164}\ } (\bibinfo {year} {2022})}\BibitemShut {NoStop}%
\bibitem [{\citenamefont {Bidzhiev}\ \emph {et~al.}(2023)\citenamefont
  {Bidzhiev}, \citenamefont {Wennersteen}, \citenamefont {Beji}, \citenamefont
  {Dagrada}, \citenamefont {D'Arcangelo}, \citenamefont {Grijalva},
  \citenamefont {Henaff}, \citenamefont {Quelle},\ and\ \citenamefont
  {Naik}}]{bidzhiev2023emutn}%
  \BibitemOpen
  \bibfield  {author} {\bibinfo {author} {\bibfnamefont {K.}~\bibnamefont
  {Bidzhiev}}, \bibinfo {author} {\bibfnamefont {A.}~\bibnamefont
  {Wennersteen}}, \bibinfo {author} {\bibfnamefont {M.}~\bibnamefont {Beji}},
  \bibinfo {author} {\bibfnamefont {M.}~\bibnamefont {Dagrada}}, \bibinfo
  {author} {\bibfnamefont {M.}~\bibnamefont {D'Arcangelo}}, \bibinfo {author}
  {\bibfnamefont {S.}~\bibnamefont {Grijalva}}, \bibinfo {author}
  {\bibfnamefont {A.-C.~L.}\ \bibnamefont {Henaff}}, \bibinfo {author}
  {\bibfnamefont {A.}~\bibnamefont {Quelle}},\ and\ \bibinfo {author}
  {\bibfnamefont {A.~S.}\ \bibnamefont {Naik}},\ }\bibfield  {title} {\bibinfo
  {title} {Cloud on-demand emulation of quantum dynamics with tensor
  networks},\ }\href {https://arxiv.org/abs/2302.05253} {\bibfield  {journal}
  {\bibinfo  {journal} {arXiv:2302.05253}\ } (\bibinfo {year}
  {2023})}\BibitemShut {NoStop}%
\bibitem [{\citenamefont {Rudolph}\ \emph {et~al.}(2022)\citenamefont
  {Rudolph}, \citenamefont {Miller}, \citenamefont {Chen}, \citenamefont
  {Acharya},\ and\ \citenamefont {Perdomo-Ortiz}}]{rudolph_synergy_2022}%
  \BibitemOpen
  \bibfield  {author} {\bibinfo {author} {\bibfnamefont {M.~S.}\ \bibnamefont
  {Rudolph}}, \bibinfo {author} {\bibfnamefont {J.}~\bibnamefont {Miller}},
  \bibinfo {author} {\bibfnamefont {J.}~\bibnamefont {Chen}}, \bibinfo {author}
  {\bibfnamefont {A.}~\bibnamefont {Acharya}},\ and\ \bibinfo {author}
  {\bibfnamefont {A.}~\bibnamefont {Perdomo-Ortiz}},\ }\bibfield  {title}
  {\bibinfo {title} {Synergy between quantum circuits and tensor networks:
  Short-cutting the race to practical quantum advantage},\ }\href
  {https://arxiv.org/abs/2208.13673} {\bibfield  {journal} {\bibinfo  {journal}
  {arXiv:2208.13673}\ } (\bibinfo {year} {2022})}\BibitemShut {NoStop}%
\bibitem [{\citenamefont {Sim}\ \emph {et~al.}(2019)\citenamefont {Sim},
  \citenamefont {Johnson},\ and\ \citenamefont
  {Aspuru-Guzik}}]{sim2019expressibility}%
  \BibitemOpen
  \bibfield  {author} {\bibinfo {author} {\bibfnamefont {S.}~\bibnamefont
  {Sim}}, \bibinfo {author} {\bibfnamefont {P.~D.}\ \bibnamefont {Johnson}},\
  and\ \bibinfo {author} {\bibfnamefont {A.}~\bibnamefont {Aspuru-Guzik}},\
  }\bibfield  {title} {\bibinfo {title} {Expressibility and entangling
  capability of parameterized quantum circuits for hybrid quantum-classical
  algorithms},\ }\href {https://doi.org/10.1002/qute.201900070} {\bibfield
  {journal} {\bibinfo  {journal} {Advanced Quantum Technologies}\ }\textbf
  {\bibinfo {volume} {2}},\ \bibinfo {pages} {1900070} (\bibinfo {year}
  {2019})}\BibitemShut {NoStop}%
\bibitem [{\citenamefont {Holmes}\ \emph {et~al.}(2022)\citenamefont {Holmes},
  \citenamefont {Sharma}, \citenamefont {Cerezo},\ and\ \citenamefont
  {Coles}}]{holmes2022connecting}%
  \BibitemOpen
  \bibfield  {author} {\bibinfo {author} {\bibfnamefont {Z.}~\bibnamefont
  {Holmes}}, \bibinfo {author} {\bibfnamefont {K.}~\bibnamefont {Sharma}},
  \bibinfo {author} {\bibfnamefont {M.}~\bibnamefont {Cerezo}},\ and\ \bibinfo
  {author} {\bibfnamefont {P.~J.}\ \bibnamefont {Coles}},\ }\bibfield  {title}
  {\bibinfo {title} {Connecting ansatz expressibility to gradient magnitudes
  and barren plateaus},\ }\href {https://doi.org/10.1103/PRXQuantum.3.010313}
  {\bibfield  {journal} {\bibinfo  {journal} {PRX Quantum}\ }\textbf {\bibinfo
  {volume} {3}},\ \bibinfo {pages} {010313} (\bibinfo {year}
  {2022})}\BibitemShut {NoStop}%
\bibitem [{\citenamefont {Tangpanitanon}\ \emph {et~al.}(2020)\citenamefont
  {Tangpanitanon}, \citenamefont {Thanasilp}, \citenamefont {Dangniam},
  \citenamefont {Lemonde},\ and\ \citenamefont
  {Angelakis}}]{tangpanitanon2020expressibility}%
  \BibitemOpen
  \bibfield  {author} {\bibinfo {author} {\bibfnamefont {J.}~\bibnamefont
  {Tangpanitanon}}, \bibinfo {author} {\bibfnamefont {S.}~\bibnamefont
  {Thanasilp}}, \bibinfo {author} {\bibfnamefont {N.}~\bibnamefont {Dangniam}},
  \bibinfo {author} {\bibfnamefont {M.-A.}\ \bibnamefont {Lemonde}},\ and\
  \bibinfo {author} {\bibfnamefont {D.~G.}\ \bibnamefont {Angelakis}},\
  }\bibfield  {title} {\bibinfo {title} {Expressibility and trainability of
  parametrized analog quantum systems for machine learning applications},\
  }\href {https://doi.org/10.1103/PhysRevResearch.2.043364} {\bibfield
  {journal} {\bibinfo  {journal} {Phys. Rev. Res.}\ }\textbf {\bibinfo {volume}
  {2}},\ \bibinfo {pages} {043364} (\bibinfo {year} {2020})}\BibitemShut
  {NoStop}%
\bibitem [{\citenamefont {de~Léséleuc}\ \emph {et~al.}(2018)\citenamefont
  {de~Léséleuc}, \citenamefont {Barredo}, \citenamefont {Lienhard},
  \citenamefont {Browaeys},\ and\ \citenamefont
  {Lahaye}}]{de_leseleuc_analysis_2018}%
  \BibitemOpen
  \bibfield  {author} {\bibinfo {author} {\bibfnamefont {S.}~\bibnamefont
  {de~Léséleuc}}, \bibinfo {author} {\bibfnamefont {D.}~\bibnamefont
  {Barredo}}, \bibinfo {author} {\bibfnamefont {V.}~\bibnamefont {Lienhard}},
  \bibinfo {author} {\bibfnamefont {A.}~\bibnamefont {Browaeys}},\ and\
  \bibinfo {author} {\bibfnamefont {T.}~\bibnamefont {Lahaye}},\ }\bibfield
  {title} {\bibinfo {title} {Analysis of imperfections in the coherent optical
  excitation of single atoms to {Rydberg} states},\ }\href
  {https://doi.org/10.1103/PhysRevA.97.053803} {\bibfield  {journal} {\bibinfo
  {journal} {Physical Review A}\ }\textbf {\bibinfo {volume} {97}},\ \bibinfo
  {pages} {053803} (\bibinfo {year} {2018})}\BibitemShut {NoStop}%
\bibitem [{\citenamefont {Henriet}(2020)}]{henriet_robustness_2020}%
  \BibitemOpen
  \bibfield  {author} {\bibinfo {author} {\bibfnamefont {L.}~\bibnamefont
  {Henriet}},\ }\bibfield  {title} {\bibinfo {title} {Robustness to spontaneous
  emission of a variational quantum algorithm},\ }\href
  {https://doi.org/10.1103/PhysRevA.101.012335} {\bibfield  {journal} {\bibinfo
   {journal} {Physical Review A}\ }\textbf {\bibinfo {volume} {101}},\ \bibinfo
  {pages} {012335} (\bibinfo {year} {2020})}\BibitemShut {NoStop}%
\bibitem [{\citenamefont {Guo}\ \emph {et~al.}(2022)\citenamefont {Guo} \emph
  {et~al.}}]{guo2022chemistry}%
  \BibitemOpen
  \bibfield  {author} {\bibinfo {author} {\bibfnamefont {S.}~\bibnamefont
  {Guo}} \emph {et~al.},\ }\bibfield  {title} {\bibinfo {title} {Scalable
  quantum computational chemistry with superconducting qubits},\ }\href
  {https://arxiv.org/abs/2212.08006} {\bibfield  {journal} {\bibinfo  {journal}
  {arXiv:2212.08006}\ } (\bibinfo {year} {2022})}\BibitemShut {NoStop}%
\bibitem [{\citenamefont {Wu}\ \emph {et~al.}(2023)\citenamefont {Wu},
  \citenamefont {Sun}, \citenamefont {Huang},\ and\ \citenamefont
  {Yuan}}]{wu_overlapped_2023}%
  \BibitemOpen
  \bibfield  {author} {\bibinfo {author} {\bibfnamefont {B.}~\bibnamefont
  {Wu}}, \bibinfo {author} {\bibfnamefont {J.}~\bibnamefont {Sun}}, \bibinfo
  {author} {\bibfnamefont {Q.}~\bibnamefont {Huang}},\ and\ \bibinfo {author}
  {\bibfnamefont {X.}~\bibnamefont {Yuan}},\ }\bibfield  {title} {\bibinfo
  {title} {Overlapped grouping measurement: {A} unified framework for measuring
  quantum states},\ }\href
  {https://quantum-journal.org/papers/q-2023-01-13-896/} {\bibfield  {journal}
  {\bibinfo  {journal} {Quantum}\ }\textbf {\bibinfo {volume} {7}},\ \bibinfo
  {pages} {896} (\bibinfo {year} {2023})}\BibitemShut {NoStop}%
\end{thebibliography}%

\end{document}